\documentclass[mathleft
]{an}
\usepackage{graphicx}
\usepackage{rotating}
\usepackage{lscape}
\usepackage{longtable}
\usepackage{times}
\begin{document}

\sloppy

\Pagespan{1}{20}
\Yearpublication{2018}%
\Yearsubmission{2017}%
\Month{11}%
\Volume{999}%
\Issue{88}%

\title{New sunspots and aurorae in the historical Chinese text corpus? \\
Comments on uncritical digital search applications}

\author{D.L. Neuh\"auser\inst{1}, R. Neuh\"auser\inst{2} \thanks{E-mail: rne@astro.uni-jena.de}, J. Chapman\inst{3,4}}

\titlerunning{Sunspots and aurorae in the historical Chinese text corpus}
\authorrunning{D.L. Neuh\"auser et al.}

\institute{
$^{1}$ Schillbachstrasse 42, 07745 Jena, Germany \\
$^{2}$ Astrophysikalisches Institut, Universit\"at Jena, Schillerg\"asschen 2-3, 07745 Jena, Germany \\
$^{3}$ Department of History, University of Oklahoma-Norman, 455 West Lindsey Street, Room 403A, Norman, Oklahoma
73019-2004, United States \\
$^{4}$ Department of History, University of California-Merced, COB2, 5200 North Lake Road, Merced, CA 95343,
United States
}

\received{July 2017}
\accepted{Oct 2017}
\publonline{ }


\abstract{
We review some applications of the method of electronic searching for historical observations of sunspots and
aurorae in the Chinese text corpus
by Hayakawa et al. (2015, 2016, 2017ab), Kawamura et al. (2016), and Tamazawa et al. (2017).
However, we show strong shortcomings in the digital search technique
as applied by them: almost all likely true sunspot and aurora
records were presented before (e.g. Xu et al. 2000), which is not mentioned in those papers;
the remaining records are dubious and often refer to other phenomena, neither spots nor aurorae
(this also applies to Hayakawa et al. 2017c).
The alleged aurorae in Hayakawa et al. (2015) and Kawamura et al. (2016) show a broad peak around full moon, not
expected for aurorae.
Hayakawa et al. (2017a) use the Korean report
\textit{At night, the gate of heaven was opened} (between AD 992 Dec 26 and 993 Jan 25, i.e. close to the $^{14}$C variation AD
993/4) to estimate the Dst index of solar activity, even though the text does not fulfil any discriminative aurora
criteria (except night-time).
Most of the above publications include very few Chinese texts and translations, and their tables with
abbreviated keywords do not allow the reader to consider alternative interpretations
(the tables also do not specify which records mention night-time). We have compared some of their event
tables with previously published catalogues and found various discrepancies. There are also intrinsic inconsistencies,
misleading information (lunar phase for day-time events), and dating errors.
We present Chinese texts and translations for some of their presumable new aurorae:
only one can be considered a likely true aurora (AD 604 Jan);
some others were selected on the sole basis of the use of the word
\textit{light} or \textit{rainbow}. Several alleged new aurorae
present observations beside the Sun during day-time.
There are well-known comets among their presumable aurorae.
We also discuss, (i) whether \textit{heiqi ri pang} can stand for black
spot(s) \textit{on one side of} or \textit{beside} the sun, (ii) aurora
color confusion in Hayakawa et al. (2015, 2016), and (iii)
whether \textit{white} and \textit{unusual rainbows} can be aurorae.
}

\maketitle

\section{Introduction}

Solar activity of the past centuries and millennia can be reconstructed with radioisotopes, sunspots, and aurorae.
Historical observations of these phenomena recently attracted new interest because of the expectation that strong $^{14}$C
variations as published for AD 775 and 994 (Miyake et al. 2012, 2013) may be related to large sunspots and strong
aurorae if these were due to super-strong solar flares. Usoskin et al. (2013) and Zhou et al. (2014) searched for
aurorae around AD 775. Stephenson (2015) searched for general astronomical evidence around AD 774/5 and concluded:
{\em Were the years around AD 774/5 ... indeed unusual for the frequency of aurorae ... ? ... both East Asian
and European records suggest that the answer seem to be in the negative ... [T]here is little sign for unusual solar
activity}. The lack of strong aurorae (and sunspot sightings) between AD 774 to 786 led Chapman et al.
(2015) and Neuh\"auser \& Neuh\"auser (2015a) to cast doubt on the flare hypothesis, so that Neuh\"auser \& Neuh\"auser (2015c)
suggested instead that solar activity dropped markedly for a few years. This suggestion is consistent with weak solar
wind and, hence, strong cosmic ray influx and radioisotope production on Earth.

Many catalogues of historical aurorae have been compiled, e.g. Matsushita (1956), Keimatsu (1970, 1971, 1972, 1973, 1974,
1975, 1976), Yau et al. (1995), Xu et al. (2000), Lee et al. (2004), and Chapman et al. (2015) for East Asian records
and de Mairan (1733), Sch\"oning (1760), Jeremiah (1870), Fritz (1873), Vyssotsky (1949), Link (1962), Rethly \& Berkes
(1963), Newton (1972), and Stothers (1979) for European records, as well as Rada \& al-Najeh (1997), Cook (2001), and
Basurah (2005, 2006) for West Asian reports; Schove (1955, 1964, 1984), Dall'Olmo (1979), Krivsky \&
Pejml (1988), Silverman (1998), Hetherington (1996), Usoskin et al. (2013), Stephenson (2015) and
Neuh\"auser \& Neuh\"auser (2015a) compiled catalogues for several cultures (partly for short time intervals). 

Naked-eye sunspots records were compiled mainly by Keimatsu (1970, 1971, 1972, 1973, 1974, 1975, 1976), Clarke \& Stephenson
(1978), Wittmann \& Xu (1987), Yau \& Stephenson (1988), Xu et al. (2000), Vaquero et al. (2002), Lee et al. (2004),
and Vaquero (2007).

All these catalogues were compiled by individually checking candidates across multiple primary sources.

Recently, Hayakawa et al. (2015, henceforth H+15) used the electronic database \textit{Scripta Sinica} of the Academia
Sinica, Taiwan, to search electronically for certain words or combinations of words referring to aurorae and sunspots
in the Chinese \textit{Song} dynasty chronicle for the period AD 960-1279. Within only the {\em Astronomic Treatise} in
the \textit{Song Shi} (History of the Song Dynasty), they searched for aurora and sunspot records, while others have
previously searched for aurorae and sunspots also in other documents covering the same period, inside and outside the
\textit{Song Shi}. Then, Hayakawa et al. (2016, henceforth H+16) extended their search for more wordings, e.g.
considering {\em white rainbows} ({\em bai ni}) as aurorae. 

In this paper, we show shortcomings in the blind electronic search technique as employed by H+15, H+16, and 
Kawamura et al. (2016, henceforth K+16)
as well
as in Tamazawa et al. (2017) and Hayakawa et al. (2017a, 2017b).

We stress that some previous other aurora catalogues are also not free of misidentifications (other phenomena), as
studied in detail for the decades around AD 775 by Chapman et al. (2015) and Neuh\"auser \& Neuh\"auser (2015ab), e.g.:

(a) Silverman (1998) has 49 entries from AD 731 to 825 in his online aurora catalogue (now at spdf.sci.gsfc.nasa.gov),
but some events are listed several times, and some of them are likely non-auroral: there are 17 different likely true
aurorae remaining in Silverman's catalogue from AD 731 to 825, according to the criteria in Neuh\"auser \& Neuh\"auser
(2015a).

(b) Usoskin et al. (2013) listed 14 to 16 different events as aurorae from AD 765 to 786, but some phenomena listed are
halos or otherwise doubtful, e.g. the European events listed for AD 773 ({\em two young men on white horses}), 774 
({\em red cross}), and 776 ({\em two inflamed shields}). These were all halo phenomena (Neuh\"auser \& Neuh\"auser 2015ab). 
Of the 14 to 16 events listed, nine are likely true or potential aurorae (Neuh\"auser \& Neuh\"auser 2015a).

(c) The global {\em Chronicle of pre-telescopic astronomy} by Hetherington (1996) includes for example the {\em red cross}
of AD 776 three times, namely in AD 773, 774, and 776, twice with wrong reference; in fact, it was most likely a halo
display in AD 776 (Newton 1972, Neuh\"auser \& Neuh\"auser 2015ab).

The often referenced aurora catalogues by Fritz (1873) and Link (1962), who partly revised Fritz (1873), likewise
include non-auroral events such as halo displays (e.g. Neuh\"auser \& Neuh\"auser 2015ab).

After discussing the criteria in the automatic searches discussed (e.g., Hayakawa et al. 2015, H+15) in Sect. 2.1, we
consider whether their electronic search technique can be considered complete (Sect. 2.2). In Sect. 3, we discuss
publications by Hayakawa et al. (2015, H+15), Kawamura et al. (2016), and Tamazawa et al. (2017) on various dynasties.
In Sect. 4, we discuss certain terminology issues of Chinese aurora and sunspot records, namely {\em ri pang} 
as in Hayakawa et al. (2017b) (Sect. 4.1), aurora color as in Hayakawa et al. (2015, 2016) (Sect.
4.2), and {\em rainbows} as in Hayakawa et al. (2016) (Sect. 4.3). We present final remarks in
our conclusion (Sect. 5).

\section{Digital searches for aurorae and sunspots}

Hayakawa et al. (2015, H+15) searched for aurorae and sunspots only in the \textit{Tianwen zhi} (Treatise on
Celestial Patterns, or more loosely, Treatise on Astronomy) in the \textit{Song Shi}. We will now discuss various
aspects of the search technique. Our considerations also apply to Hayakawa et al. (2016, 2017a, 2017b), Kawamura et al.
(2016), and Tamazawa et al. (2017).

\subsection{Selection criteria}

H+15 searched for {\em descriptions that could be regarded as records of sunspots or auroras {\dots} such as black spot and
red vapor} (their section {\em search method}). H+15 add that {\em sunspots are described as black spots or black vapors in
the sun or in terms such as ``the sun was weak and without light''} (their section on {\em sunspot records}). According to
their table 1 on sunspots, they also included a report on the {\em crescent-like shape} (of the sun) as a report on
sunspot(s) (AD 1005 Feb 6). In their section on {\em auroral records}, they add that they {\em surveyed the words that refer
to luminous phenomena, such as vapor, light, and cloud}, excluding those without dates and those observed explicitly
during the day (H+15); H+15 also 
flag 
{\em red sunsets} and {\em red sunrises} as aurorae (their table 2). K+16
mention that they searched for {\em vapor} ({\em qi}), {\em cloud} ({\em yun}), and {\em light} ({\rm guang}) 
as aurorae (and for {\em black spot} ({\em heizi}) and {\em black vapors} ({\em heiqi}) for sunspots, but
found none), and that they excluded those {\em associated with the Sun or Moon}, and those
explicitly observed at day-time (their section 2.1).

It would be best, if the authors would give a clear and full list of all words and combinations they have searched for.
If some texts, including certain wordings which could have been regarded as the relevant phenomenon, are not
interpreted as that particular phenomenon, one should then compile a list of false positives (i.e. other events) with
clear reasoning as to why they are dubious. 
Reports on {\em light} or {\em cloud} could 
have been anything including meteors, bolides, halo displays, comets, 
rainbows, fogbows, novae, supernovae, or other meteorological events.

Chapman et al. (2015) discussed the Chinese wording for aurorae and other celestial phenomena in more detail: "Any
study of aurorae in medieval China is complicated by the fact that there was no discrete concept of aurorae as such in
medieval Chinese astronomy. Scholars have variously identified observations of flowing stars (\textit{liu xing}) or
stars that fall (\textit{xing yun}), various sorts of halos (\textit{huan}), and \textit{qi} as aurorae. Flowing stars
and stars that fall in almost every case should be identified with meteors or bolides, while \textit{huan} (rings) are
most likely lunar or solar halo displays; flowing stars in some instances may also refer to comets. The most likely
instances of aurorae in Chinese historical records are identified as \textit{qi}, yet not all or even most observations
of \textit{qi} were indeed aurorae. While \textit{qi} is variously translated as {\em ether(s)}
or {\em vapour(s)}, material objects, including clouds, planets, stars, comets, and meteors,
were thought to be constituted of \textit{qi} in Chinese cosmology. Because \textit{qi} was thought to emanate from the
Earth itself, often in response to developments in the politico-religious sphere of the imperial court, explanations
for aberrant astronomical and meteorological phenomena were grounded in politics. {\dots} In terms of practical
observation of celestial patterns (i.e. \textit{tianwen}, a term often imprecisely translated as astronomy or
astrology), clouds (\textit{yun}) and \textit{qi} were included in a single category (\textit{yunqi}), and recorded in
the same section of the astronomical treatises."

Pre-modern scholars in Europe and Arabia also considered all transient celestial phenomena like comets and (super-)novae
(as well as aurorae) as happening in the Earth's atmosphere, following Aristotle's \textit{Meteorology}.

Furthermore, the meaning of words changes with time (usually, they get more specific). E.g., the words \textit{cometes}
in Latin and \textit{nayzak} in Arabic is mostly translated to {\em comet}; before about AD 1600, it did not only mean
{\em comet} in today's sense, but had a more general meaning of {\em apparently
extended transient celestial phenomenon} including novae and supernovae (Goldstein 1965, Stephenson \&
Green 2002, Neuh\"auser et al. 2016), comparable to {\em guest star} (\textit{kexing}) in the Chinese. We will discuss
another example below ({\em rainbow}, see Sect. 4.3).

In Neuh\"auser \& Neuh\"auser (2015a), criteria are suggested which indicate likely true aurora borealis; such
characteristics are indeed needed given the heterogeneous phenomena and reports: \\
(1) Sky brightness: Night-time, whereas day-time excludes an auroral interpretation; sightings at twilight do not
fulfill the night-time criterion, but have to be investigated in more detail (sometimes, night-time is indicated
indirectly, e.g. by mentioning stars or constellations). The lunar phase should also be considered (aurorae are less
likely to be noticed if the sky is too bright).\\
(2) Direction: aurorae borealis, as their Latin name implies, are normally in the northern region; wordings like
{\rm rays to/from the zenith}, {\em east to west}, and
{\em west to east} are also well possible. While aurorae borealis can also partly be seen to
the south, if the observer is located north of the aurora oval, a southern direction
usually contradicts an auroral interpretation; sightings in the south do not fulfil the direction criterion. \\
(3) Color: phenomena with reddish, fiery, blood(y), scarlet, purple, crimson, and similar color can be auroral,
sometimes reported together with green, blue, yellow, or violet,
while blue, yellow, violet, or {\em black/dark} alone without reddish would be dubious, 
because one should expect the mention of some red in addition to other colors;
green aurorae are seen only far north;
reports on other phenomena (e.g. white, bright, brilliant, glow, light) have to be investigated 
in more detail (but do not fulfil an aurora criterion). \\
(4) Dynamics like pulses, changes, motion, etc.: the words \textit{fiery} and \textit{fight} can indicate dynamics; 
the word \textit{fire} seems to indicate dynamics only in (heterogeneous) European reports, while \textit{fire} in (more
homogeneous) East Asian reports only indicates a reddish color; to discriminate them from other variable phenomena like
halo displays, a careful analysis of text and context is needed. \\
(5) Repetition: aurorae may occur in the following nights, but not for longer than a few days. \\
See Neuh\"auser \& Neuh\"auser (2015a) for more details.

The aim of such criteria is mainly to differentiate the categorization of a report from other possible celestial events.
Even if several criteria are fulfilled, one still has to check whether the whole report is more likely pointing to an
aurora rather than some other phenomenon: e.g. a report like {\em white vapor moved in the north for several
nights} would fulfil four criteria (night, north, motion, repetition), but can mean a comet (moving relative to the
stars from night to night), or even a certain lunar halo phenomenon or a lunar fog bow (opposite of the moon and moving
with the moon during each night, certain weather situations can support halo displays for several nights in a row);
if it referred to an aurora, some mention of a color should be expected.
While the number of criteria fulfilled gives some indication as to the
likelihood that the event was an aurora, it does not quantify its strength. However, when using the aurora reports to
investigate solar activity strength or to reconstruct the Schwabe cycle, using the more likely events (with more
criteria fulfilled) is sufficient, while events with low or zero likelihood (e.g. {\em white bands} or {\em white
rainbows}) are not helpful (see Neuh\"auser \& Neuh\"auser 2015a for examples). 

H+15, H+16, and K+16 do not discuss any criteria in sufficient detail. They largely avoid discussion of individual
events or alternative interpretations of the records.

Furthermore, while H+15, H+16, and K+16 do exclude observations explicitly reported for full daylight, they list observations at
twilight, while Neuh\"auser \& Neuh\"auser (2015a) and Chapman et al. (2015) calculated during which twilight phase (civil,
nautical, or astronomical) the sighting occurred at the given location and time. H+15 and K+16 do not indicate in their
event tables, which observations were reported for night-time. Tamazawa et al. (2017) also claim to leave
out day-time observations in their aurora list, but they do list many day-time sightings among their aurorae (our Sect.
3.3).

For likely true 
aurorae, 
quasi-simultaneous observations 
from other parts of the world provide additional information --
this was
not done in H+15, but, e.g., in Willis et al. (2005) or Neuh\"auser \& Neuh\"auser (2015a),
and to a limited extent in K+16, but only with the Fritz (1873) catalogue.

H+15, H+16, and K+16 provide only a few examples with Chinese text and full English translations. A list of
Chinese texts is given only on their web page, but without English translations, while full original Chinese texts
with full English translations are found elsewhere, e.g., in Keimatsu (1970, 1971, 1972, 1973, 1974, 1975, 1976), Xu et al.
(2000), and Chapman et al. (2015).

It is of course useful to consider the lunar phase;
however, giving a lunar phase for day-time observation, as e.g. in Tamazawa et al. (2017), is misleading.
A peak of presumable aurorae around full moon (e.g. figure 3 in
K+16) shows that many non-auroral events are considered, so that more precise categorization should have been applied. 

\subsection{Completeness}

The official Chinese dynasty reports on celestial phenomena are based on well-trained court astronomers, who use a
similar protocol for reporting their observations, so that these records are more homogeneous than historical sources
from other parts of the world, in particular Europe. 
From the fact that not all the solar eclipses, which could have been observed at the
respective Chinese capital, are listed, it is clear that these compilations are not complete for day-time solar
observations; and from the fact that not all the (observable) lunar eclipses are listed, it is also evident that the
night-time observations are not complete -- due to lost documents, overcast sky conditions, varying interest in
certain phenomena, and maybe other biases (e.g. that the interpretation of an event was negative/unfortunate, so that
the reporting was omitted/avoided for political or other reasons), so that it is always worth checking also
other documents such as local records. The \textit{Qing} dynasty source studied by K+16 does include local
observations outside the capital -- still, K+16 notice (their section 3.3) that aurorae are completely missing for AD
1771-1814 (partly during the three very intense Schwabe 
cycles
no. 2 to 4 at the end of the 19th century, partly already
in the Dalton minimum). K+16 did not find any naked-eye sunspot record in the \textit{Qing} dynasty source, even
though 44 Chinese naked-eye sunspots were previously published by others (e.g. Yau \& Stephenson 1988, Xu et al. 2000)
for their study period (the time AD 1559-1912 according to their abstract), and also no sunspot nor aurora records
corresponding to the Carrington event AD 1859 were found in K+16.

There are two more reasons to study not only electronic copies of dynastic compilations: The records available
nowadays are in most cases not the original manuscripts, but copies of copies of copies. It is
well possible that scribal errors are hidden in the texts, and also dating errors (see, e.g., Chapman et al. 2015 for
examples). In a digital search it may therefore easily be possible to miss a sunspot or aurora only due to a small
scribal error, e.g. one transposed or missing letter or sign. In addition, local records often offer complimentary
information to solve obvious scribal and dating errors.

In both old manuscripts and copies the dating can be off by anything between a day and years (e.g. Newton 1972).
Manuscripts can have a known constant offset in years for a certain range of years (e.g. some versions of the
Anglo-Saxon Chronicle around AD 774, see Neuh\"auser \& Neuh\"auser 2015b). East Asian chronicles like those studied by
H+15, H+16, and K+16 have the advantage that dates are often given in two systems, both on the lunar calendar (lunar
month starting with conjunction of Moon and Sun as true new moon) and the day number in their sexagenary cycle
(counting the days from 1 to 60, then starting over again and again). In some cases, when, e.g., a sexagenary day is
given which did not exist in the given lunar month, it is likely that the lunar month (or year) is off by usually one
month (or year), or there is a mistake in the sexagenary day, e.g. a scribal error (see Chapman et al. 2015). While
e.g. Yau \& Stephenson (1988) and Xu et al. (2000) notice such cases and suggest corrections, H+15, H+16, and K+16 do
not discuss such problems individually, but instead seem to always give the given month (or year) only without
day/date, instead of correcting the date. They also neither cite nor discuss the corrections in scholarly publications.
See below for examples.

Also, studying whether there are quasi-simultaneous observations of both sunspots and aurorae within a few days is
useful to cross-check the reliability of such records (they are often related to each other) including the dating
accuracy, as e.g. done by Willis et al. (2005). H+15 would have had the possibility to check for quasi-simultaneous
observations of both phenomena easily in their database, as they list both sunspots and aurorae. There are two such
instances in the study by Willis et al. (2005) within the time period studied by H+15, namely AD 1137 Mar and AD 1193
Dec. For these dates, both the spots and the aurorae are listed by H+15 as having been seen within a few days of one
another, but it is not mentioned that the events were quasi-simultaneous. While K+16 did not find naked-eye sunspots
in the \textit{Qing} dynasty source (early 17th to early 20th century), it would have been possible to compare their
presumable aurorae to (well-dated) telescopic sunspot observations; K+16 do search for simultaneous aurora observations
outside China, but only with the catalogue by Fritz (1873), which also does include non-auroral events.

Hence, limiting a study to the official treatises and a blind electronic search significantly limits the results.

Hayakawa et al. (2017b) argue that {\em extending our survey to these [other, local, etc.] records ... can
easily overestimate the actual amount of observations ... including other kinds of records without any criteria can
make us overrate or underrate solar activity by this bias how many historical sources are still available
today} (their section on {\em Historical documents for additional discussion}).
Apart from the fact that one should not search {\em without any criteria}, the official
histories have different grades of completeness for different epochs and dynasties, and extending the survey to other
documents alone cannot overestimate solar activity, but in all such studies, it is necessary to consider the
completeness of the records, e.g. by comparing to the completeness of comet or eclipse reports (e.g. Strom 2015, Neuh\"auser et
al., in prep.).

\section{Comments on recent searches for aurorae and sunspots from different dynasties}

In this section, we comment on recent uncritical applications of the automatic search technique for aurorae and sunspots in
the Chinese text corpus.

\subsection{Song dynasty AD 960-1279 (Hayakawa et al. 2015)}

First, we will discuss sunspots, then aurorae as presented by Hayakawa et al. (2015, henceforth H+15).

\begin{figure*}
{\includegraphics[angle=0,width=17cm]{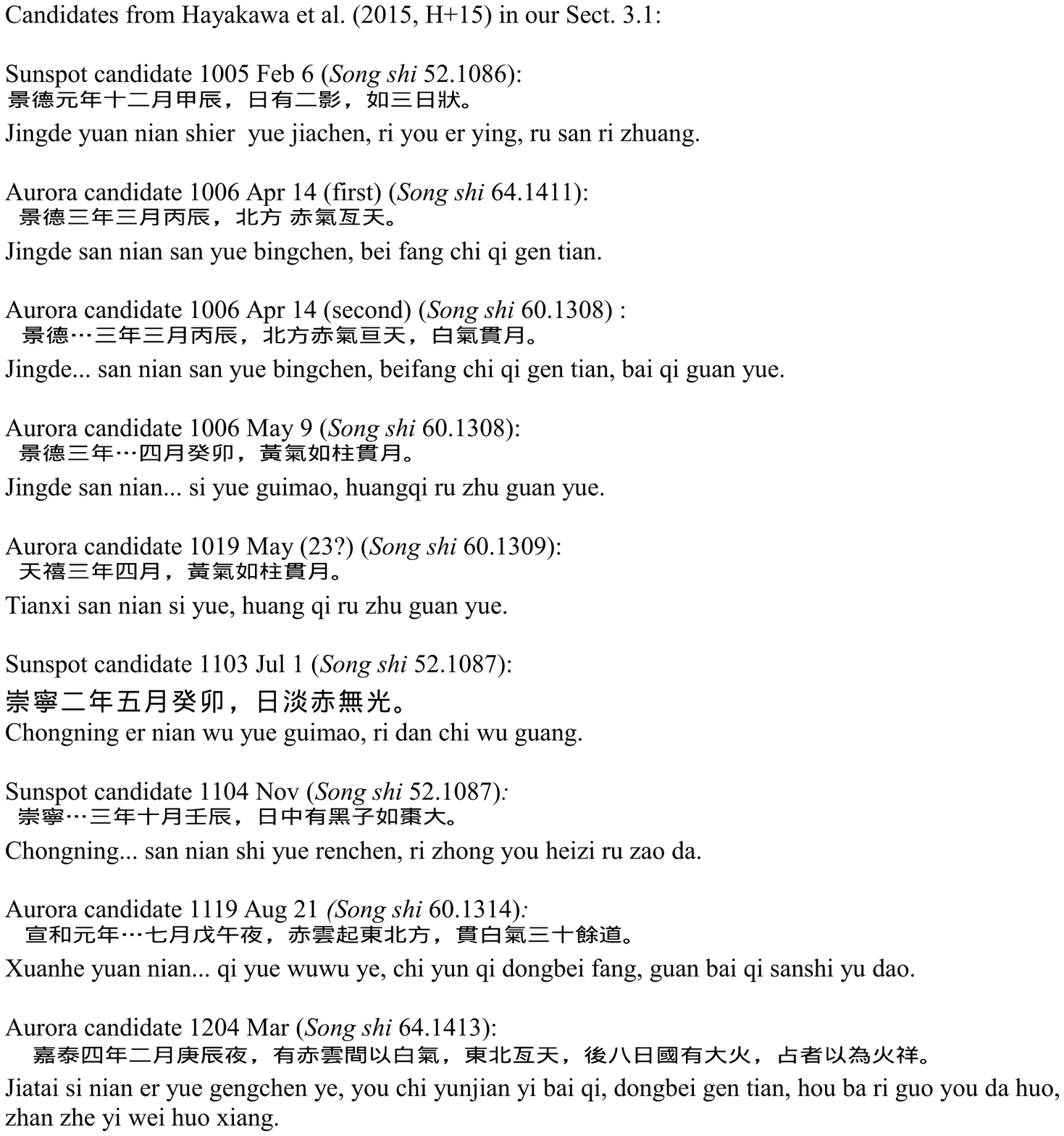}}
\caption{Here, we show the Chinese texts related to some of those events reported by Hayakawa et al. (2015),
which we discuss in our text. Translation and discussion are given in the Sect. 3.1.}
\end{figure*}

\subsubsection{Sunspot records in Hayakawa et al. (2015)}

There are inconsistencies between the two sunspot quotations (a) and (b) in H+15, as quoted in their section on
{\em sunspot records}, and their own table 1 (on sunspots): 

Instead of a spot for AD 1077 Jan 11 (H+15 quotation (a): {\em in the sun were black spots ... They disappeared on 22},
(i.e. plural {\em spots}), there is an entry for a spot in their table 1 for AD 1079 Jan 11 for 12 days (i.e. Jan
11-22) with no entry in the column on {\em counts} (indicating one spot); Xu et al. (2000) list two similar
entries from \textit{Song Shi}, use singular {\em spot} in their translation, and give AD 1079 Jan 11-22 as date range
(but none in AD 1077 Jan).

Instead of a spot on AD 1145 June (H+15 quotation (b) without day/date: {\em in the sun were black vapors}, i.e.
plural), there is an entry in their table 1 for AD 1145 July (without days/dates) with {\em black vapor} (in addition to
another entry for AD 1145 July for {\em black spot}), again without entry in the column of
{\em counts}. Yau \& Stephenson (1988) mention a possible scribal error as reason for the date problem in AD 1145
June/July; Xu et al. (2000) then suggest July 23 as date for {\em black vapor}, for which both Yau \& Stephenson (1988)
and Xu et al. (2000) use singular. While the two latter (older) publications indicate that this spot was seen for two
days, H+15 does not mention this duration (neither in the quotation nor in their table 1), even though it is found in
the \textit{Song Shi}, their source.

H+15 should have compared their sunspots and their translations with previous publications, e.g. Keimatsu (1974, 1975,
1976), Wittmann \& Xu (1987), Yau \& Stephenson (1988), Xu et al. (2000). We note the following differences to Xu et
al. (2000), the latest compilation:

While H+16 list a record for {\em 974 March} without day/date, other publications give 974 Mar 3 as date (the \textit{Song
Shi} incorrectly gives 1st month, but it was the 2nd month, e.g. Xu et al. 2000).

For AD 1005 Feb 6, H+15 give {\em shade}, {\em Counts: 2}, and
{\em crescent-like shape} (their table 1). The original Chinese text (given only on their web
page) was translated by us to: {\em In the sun there were two shades, shaped like crescents} (Chinese text in our Fig.
1). This wording was otherwise not used for likely true sunspots. Hirayama (1889) gave {\em fleckles} 
for AD 1005 Jan 10 (no source given), which may be the same observation, but dated about one
lunar month earlier; 
on AD 1005 Jan 13, there
was a total solar eclipse visible in China, so that what was interpreted as spots
by H+15 and Hirayama (1889) may be a corrupt report of an eclipse.

For AD 1079, H+15 list two records from the \textit{Song Shi}, one for Jan 11-22 and one for Mar 20 only, both
presenting a {\em plum}, while others list the latter one to last until Mar 29 based on the \textit{Song Shi} (Xu et al. 2000).

The record for AD 1103 Jul 1 reads {\em The sun was weak and without light} (listed in H+15 in and below table 1), so that
it could indicate haze or dust due to a volcanic eruption or a period of somewhat overcast days during which the sun
barely appears through a thin layer of cloud.
We found the entry for this event in \textit{Song Shi} 52.1087 
(Chinese text in our Fig. 1): 
{\em On the guimao (40) day in the fifth month of
the second year of the Chongning reign period (1 Jul 1103), the sun was faint red and without light}; the word
{\em red} was omitted in the H+15 translation; this record is not listed in other recent compilations
of sunspots.

The record for AD 1104 Nov ({\em renchen day ... within the sun there was a black spot, as large as a jujube}, i.e. a
date, our translation of the Chinese text in Fig. 1) is also not listed in any other catalogues for that date. It is
based on \textit{Song Shi wu} ch. 52, where the date is incorrectly given as {\em Emperor Huizong of Song, 3rd year of the
Chongning reign period, 10th month, day ren-chen (29)}, which should be {\em 4th year} as in
\textit{Song shi yi} ch. 20, as specified in Xu et al. (2000), there was no renchen day (29) in the 10th month of the
3rd year. Hence, the correct date is AD 1105 Dec 6 (Xu et al.: {\em There was a black spot in the sun as large as a
date}). Based on the above mentioned record from {\it Song shi yi} ch. 20, H+15 uncritically adopt and list the date given
there, too, namely AD 1105 Nov 6 (as in Keimatsu 1975), but the date has to be corrected to AD 1105 Dec 6; see e.g. Xu
et al. (2000).

The spot given for AD 1137 Mar 11 only (H+15) was observed from Mar 1-11, e.g. Xu et al. (2000).

In AD 1137, H+15 give a record for May 8 {\em until 22}; the translation in Xu et al. (2000)
dated AD 1137 May 8 reads: {\em There was a black spot on the Sun that lasted through the 5th month, when it
dissipated}, the 5th month being May 22-Jun 19 according to Xu et al. (2000). It may not be justified to
limit the visibility of the spot to May 22 (H+15), when the 5th lunar month just started, as it may be unlikely to
detect a spot on 15 subsequent days by naked eye; at least one more (later) spot could be involved given the wording
{\em through the 5th month}.

The event listed by H+15 dated {\em 1139 Mar {\dots} for a month} can better be dated {\em Mar 3 to Apr 1} 
and was observed for {\em more than a month} (e.g. Xu et al. 2000). 

The two entries for AD 1145 July in H+15 (without day/dates), one on a black vapour and one on a black spot, were dated
July 23 and 24, respectively, in, e.g., Xu et al. (2000), who remarks that what is given as {\em 6th month} in the
\textit{Song Shi} should be {\em 7th month}; it is probably this correction that was not noticed by
H+15, so that they did not give a day at all.

For AD 1185, H+15 list one spot for Feb 10, one from Feb 15 to 27 (also in Xu et al. 2000), and one more for Feb 27, but
do not mention that a spot was also seen on Feb 11 in Korea (Lee et al. 2004).

For AD 1186, H+15 list one spot each for both May 23 and 26 for one day each, while Xu et al. (2000) show that the spot
lasted from Mar 23 through Mar 27.

Almost all the sunspot records listed by H+15 (their table 1) were presented before by others as sunspots; relevant
literature such as Xu et al. (2000) is not cited in the paper. H+15 provide the original Chinese only on their web
page, without English translation. The new records are questionable and probably due to other phenomena, not spots,
e.g. {\em sun was weak and without light}.

\subsubsection{Aurora records in Hayakawa et al. (2015)}

First, we consider aurora examples highlighted in H+15, then discuss some general problems of their work related to
aurorae, and finally touch briefly aurorae around AD 993/4 (Hayakawa et al. 2017a). 

In their example quotation (a) for aurorae, listed in their section on {\em methods}, H+15 present an aurora with a
different translation compared to previous catalogues:

H+15 translate \textit{Song shi} (\textit{Five Phases} II b, p. 1413) as follows: \\
{\em on -- March CE 1204 at night, red clouds appeared within white vapors, crossing the sky from the east to the west.
After that, conflagrations occupied the country for eight days. Thus, astrologers regarded this as a symbol of fire.}

While the same Chinese text was interpreted as aurora before by others, their translations differ. Yau et al. (1995)
dated the event to AD 1204 Mar 29 and translated (brackets their additions): \\
(i) Chia-t'ai reign period, 4th year, 2nd month, day keng-shen (57). {\em At night,
a red vapour extended across the sky} (\textit{Song Shi} 38) and \\
(ii) Chia-t'ai reign period, 4th year, 2nd month, day keng-ch'en (17) (should read
keng-shen as in preceding entry). {\em At night, a red vapour mixed with a white vapour was stretching across the N sky}
(\textit{Song Shi} 64).

Xu et al. (2000) translate the event dated AD 1204 Mar 29 as follows (square brackets their addition): \\
Emperor Ningzong of Song, 4th year of the Jiatai reign period, 2nd month, day
gengshen [57]. {\em During the night, there was a white vapor among the scarlet clouds that extended across the northeast
sky} (\textit{Wenxian tongkao} ch. 298).

The text fulfils three aurora criteria
from Neuh\"auser \& Neuh\"auser (2015a): night-time, partly northern direction, and red color. AD 1204 Mar 29 is only a few
days before the new moon of Apr 2.

We see that the older translation used the traditional Wade-Giles system for romanizing Chinese characters. More
significantly, we also see that both Yau et al. (1995) and Xu et al. (2000) date the event to AD 1204 Mar 29, while
H+15 only give {\em 1204 March} in their quotation (a). 

While H+15 translate \textit{Song shi} 64.1413 {\em from the east to the west}, Yau et al.
(record ii) give {\em across the north sky}. The original Chinese text (see our Fig. 1) does
not contain the phrase {\em from the east to the west}, H+15 misread \textit{north} for
\textit{west}, the specified direction is NE. The entry mentions red \textit{yun} (cloud) and white \textit{qi}
(vapor), both for the gengchen night in the 2nd month, i.e. Mar 29.

Quotation (a) in H+15 above gives {\em east to the west} as direction, while their own table 2 gives {\em east-north} for
this event (1204 Mar without day), the direction from Xu et al. In their table 2, there is another event dated AD 1204
Mar 29 ({\em red vapour}) without direction, obviously a duplication (see Yau et al. record i). Keimatsu (1976) also
listed this event for AD 1204 Mar 29 (as certain aurora), but without translation (published by Fukushima after
Keimatsu had passed away). H+15 even goes as far as claiming that this particular aurora (dated AD 1204 Mar without
day/date) lasted for eight days (their table 2), even though this is not specified in the record; the record in fact
states that afterwards {\em conflagrations occupied the country for eight days}. 

In their example quotation (b) for aurorae (AD 1119 Aug 21), listed in their section on {\em auroral records}, H+15
present an event where their English translation does not include the word {\em night}, even though the Chinese text (our
Fig. 1) does include the word for night. Keimatsu (1975), Yau et al. (1995), and Xu et al. (2000) all list this event
and do include its night-time occurrence in their translations. Table 2 in H+15 lists only {\em north} for this event,
while the quotation (b) in their section on {\em auroral records} gives {\em northeast}. Keimatsu (1975), Yau et al.
(1995), and Xu et al. (2000) all give {\em northeast}. The translation in H+15 says: {\em red clouds appear [qi] in the NE
direction running through 30 ways of white vapors} (at night). Other translations of the same Chinese entry (Keimatsu
1975, Yau et al. 1995, Xu et al. 2000) used {\em rose} or {\em arose}
correctly for the verb {\em qi} instead of {\em appear} (which would be
{\em xian}). This fulfils the aurora criteria night, partly north, and red, but it happened
shortly before full moon (new moons were AD 1119 Aug 8 and Sep 6, Gautschy 2014). As the moon rose that night in the
ESE as seen from Kaifeng, the northern Song capital, a lunar halo display is less likely. 

Classical Chinese often does not distinguish between passive and active senses of a verb, as is the case with
{\em penetrated} here (translation from Xu et al. 2000 instead of {\em running through} as in
H+15). The lines in question might be rendered in two ways, as is the case above. We suspect that the {\em more than
thirty `ways'} refers to shafts of light (\textit{qi} can refer to light; the convenient translation {\em vapors} may
well be leading us astray here). {\em Way} is a {\em dao}, which means in its basic sense a road, and is here used
to denote something long and thin. {\em Penetrated} should probably be understood in the passive, as something long and
thin can penetrate a cloud, while it is hard to imagine a cloud penetrating something long and thin. We therefore
translate: {\em On the night of the wuwu day in the seventh month, red clouds arose in the northeastern direction
and were penetrated by more than thirty shafts of white qi} ({\em qi} here seems to mean {\em light}).

In their table 2, H+15 provide only a few keywords from their English translations, but not the full text; the original
Chinese is given only on their web page, without English translations nor references to previous work. It would have
been beyond the scope of our paper to compare the translations by H+15 or K+16 (the keywords in their
tables) with all previously published translations, e.g. by Matsushita (1956), Keimatsu
(1970, 1971, 1972, 1973, 1974, 1975, 1976), Yau et al. (1995), or Xu et al. (2000), such a comparison should have been done
by H+15 and K+16. 

Duplications are also a problem: while H+15 list two aurorae with the same details for AD 979 May 9 and 19 (their table
2), Yau et al. (1995) mentioned that it is one event on either May 9 or 19. Other duplications are 986 Feb 22 plus one
in 986 Feb (without day), both red vapor in Kaifeng. The event listed for 1007 May 13 reports a band of white vapor
covering the moon (probably a halo phenomenon).\footnote{While Hayakawa et al. (2017b) write about their aurora YS\#A1
that {\em Yau et al. (1995) and Xu et al. (2000), with its date as 1262.02.09 mistakenly, and we should
correct this to the same date in 1261} (end of their section 3), we would like to note that Xu et al.
(2000) dated this entry to AD 1261 Feb 9. 
Abbott \& Juhl (2016) also list this event as aurora for AD 1261 Feb 9, and comment {\em N[not] I[n] Y[au]},
while it is in Yau et al. (1995) under AD 1262 Feb 9.
}

Table 2 in H+15 on aurorae implies that all events reported would be {\em auroral records}, but
in fact they are somewhat luminous, partly colorful events at night or around twilight (dusk or dawn), not all such
events are necessarily aurorae. 

We will now consider four more presumable aurorae listed in table 2 of H+15 (Chinese text in our Fig. 1), 
somewhat related to each other, which were
previously discussed in Neuh\"auser et al. (2017): there are two entries for AD 1006 Apr 14 (brackets are our additions,
but as meant in H+15):
{\em 1006 Apr 14 R[ed] V[apour] n[orth] [in] Kaifeng [lunar phase] 0.46} and {\em 1006 Apr 14 W[hite] V[apour] near the
moon [in] Kaifeng [lunar phase] 0.46} (0.46 means near full moon). One of the two texts is from the astronomical
treatise (\textit{Tianwen zhi}) of the \textit{Song Shi}, the other from its treatise on \textit{general omenology}
(\textit{wuxing zhi}). Xu et al. (2000) combine red and white vapor to one text and then also give Song shi Tianwen zhi
60 and wuxing zhi 64. The two original sources read: \\
{\em On the bingchen day of the third month of the third year of the Jingde reign period, northern direction,
red qi spread across the sky} (\textit{Song shi} 64.1411) and \\
{\em On the bingchen day of the third month of the third year of the Jingde reign period, northern direction, red qi
spread across the sky. White qi penetrated the moon.} (\textit{Song shi} 60.1308).

It is very unlikely that the white vapour near the almost full moon is an aurora (more likely some halo effect). If the
red vapour was far away from the moon or when the moon was below horizon, then it could have been an aurora.

There is an additional entry in table 2 of H+15: {\em 1006 May [without day] Y[ellow] V[apour] near the moon [in]
Kaifeng}, also from \textit{Song Shi} 60.1308; our
translation of this entry in AD 1006 is: {\em On this date (guimao, 40) yellow qi like a pillar penetrated the
moon}; hence, this is also not an aurora. The date for this event is uncertain: as it is given as the \textit{guimao}
(40) day in the fourth month, when in fact there was no \textit{guimao} day in the fourth month; a \textit{guimao} day
did occur at the beginning of the fifth month (1006 May 31) and at the beginning of the third month (1006 Apr 1). Both
are, however, so close to new moon that the text (penetrated the moon) does not fit to the given sexagenary date. There
is another instance of the same phrase ({\em yellow vapour like a pillar penetrated the moon}) dated to the 4th month of
the 3rd year of the \textit{Tianxi} reign period (AD 1019) given without the \textit{guimao} date -- even though there is a
\textit{guimao} date in that month (AD 1019 May 23, close to full moon on May 21/22). It is possible that the event
somehow got transposed to the wrong reign period (Neuh\"auser et al. 2017).

H+15 list {\em Y[ellow] V[apour]} for 1019 May 8. The text does not specify a particular day; May 8 is the first day of
the month.  They omitted the wording {\em penetrated the moon}, which is clearly given in the original Chinese (see our
Fig. 1); H+15 give 0.04 as lunar phase (new moon May 7), so that again the text (penetrated the moon) is not consistent
with the lunar phase for the date given in H+15. This event was most probably a lunar halo pillar rather than an aurora
(Neuh\"auser et al. 2017). 

Of the four entries in H+15 for AD 1006 Apr 14 red, AD 1006 Apr 14 white, AD 1006 May yellow, and AD 1019 May 8 yellow,
only the first may be an aurora. The remaining three are given as moon-related phenomena. Moreover, the last two have
dating problems and were probably a single event (wrong date in H+15).

The Chinese linear measurements \textit{cun}, \textit{chi}, and \textit{zhang} are often used to describe the length of
celestial objects, such as comets tails. They also tend be used for approximations and are frequently prefaced with the
word \textit{ke} (about). H+15, H+16, and K+16 wrote that it is unknown which angle in the sky these
measurements refer to. However, this is well known, as most recently given in Chapman et al. (2015) and references
therein: A \textit{chi} is one degree when referring to astronomical objects (angle in the sky), see Ho Peng
Yoke (1962); otherwise for terrestrial linear measures it is one foot or 25 cm (but varying somewhat in time, e.g. 33
cm in \textit{Tang} times). A \textit{zhang} is either ten Chinese feet for general linear measurements, or ten degrees
as angle on sky in astronomy (Wilkinson 2000; Wang 2008). A \textit{du} is a Chinese degree, where our (i.e. the
Babylonian) 360.00 degrees correspond to 365.25 Chinese degree (\textit{du}), one \textit{du} is then 0.98562 degree. 

The presumable aurorae plotted in figure 3 in H+15 do not correspond to the aurora dates listed in their own table 2:
e.g., the first plotted aurora is missing in their table, while other early aurorae listed in the table are not
plotted. The caption to their figure 3 speaks about {\em 774}, while the figure covers a few
decades around AD 1000.

Figure 4 in H+15 as well as figures 4 and 5 in Hayakawa et al. (2017b) show the distribution of their presumable aurorae
with lunar phase. One would expect a broad minimum around full moon (phase 0.5), because aurorae can better be found in
dark nights. These figures do not show such a drop, which indicates that there are probably lunar-related phenomena
in their data set (seen around full moon). It also is not useful to correlate a sample with twilight observations with the lunar phase:
e.g., in H+15, at least 25 of the 193 entries in their table 2 are during sunrise, sunset, or dusk. 

The period studied in H+15 starts with AD 960 and therefore includes the C14 variation around AD 994. They write that
they {\em did not find any candidates of sunspot nor aurora in these years [AD 993-994]} and that they
{\em found a cluster of auroral candidates several years after this event. The closest aurora recorded is in
996 and there is a record of a sunspot in CE 1005} (H+15 for {\textit Song shi}). Then, Hayakawa et al. (2017a) re-investigated
possible aurorae from AD 990 to 994. They argue that no Chinese observations of aurorae from AD 988 to 996 could be due
to bad weather or lost records; however, we note that there were several lunar eclipses observed in China from AD 991
to 995 (e.g. Xu et al. 2000) plus comet Halley in AD 989 with many Chinese records (e.g. Ho Peng Yoke 1962), so that celestial
observations are extant. 

Hayakawa et al. (2017a) found in Fritz (1873) and Link (1962) a few European sightings (which may be likely true
aurorae) and also an event in Korean records between 992 Dec 26 and 993 Jan 25 (12th month in
11th year of Seongjong): {\em At night, heaven's gate opened} (their section
3.4). Only the night-time criterion is fulfilled; the date range for the relevant lunar month was incorrectly given as
992 Dec 27 to 993 Jan 15 in Hayakawa et al. (2017a), four times in their section 4 (the correct date range is 992 Dec
26 and 993 Jan 25). The event listed for AD 992 Dec (without day) as {\em R[ed]} in Lee et al.
(2004) from Korea may be meant to be the same record, but the text does not give any color.
While Yau et al. (1995) still included it in their aurora catalogue, Stephenson (2015) re-considered whether
this Korean sighting could be auroral: {\em somewhat vague Korean allusion around the end of 992 or the
start of 993} (his section 5.3).

A drawing of a presumably similar event reported as {\em heaven's split} (translation in Hayakawa et al. 2017a) in a
Chinese divination manual shown in their figure 3 (Hayakawa et al. 2017a) does not support an auroral interpretation,
partly because the drawing is fully consistent with a halo phenomenon or a fog bow: a white band or bow seems to touch
the landscape (like a typical fog bow), and the text accompanying it also does not present any aurora-typical wordings
such as \textit{qi} (vapor) nor other aurora criteria. As the surrounding text depicted in Hayakawa et al. (2017a)
reveals, the \textit{Goryeosa} (\textit{History of Goryo}) astronomical treatise in which the event is recorded is
highly variegated. The record stating that {\em heaven's gate opened} occurs alongside entries for comets, meteors,
planetary conjunctions, eclipses, and other phenomena. The absence of divisions between various types of events makes
it particularly difficult to deduce what the phrase {\em heaven's gate opened} meant or what sort of phenomenon it
described. 

Hayakawa et al. (2017a) conclude from the equatorward boundary of the aurora oval (i.e. from the dubious Korean
sighting) on the strength of the geomagnetic activity (Dst index): {\em Although the estimated Dst is a crude
approximation and a careful diagnosis necessary, there is a possibility that the magnetic storm that occurred in the
period between 992.12.27 and 993.01.15 was a stronger storm than any of the storms recorded since 1957}
(Hayakawa et al. 2017a, their section 4.2), with one of the incorrectly given month ranges.
Hayakawa et al. (2017a) then speculated whether there was a series of several large coronal mass ejections
within one month. There were in fact several likely true aurorae in AD 992 Apr, Oct, and Dec in northern and central
Europe, indicating a solar activity level which is not exceptional (see also Stephenson 2015 and Neuh\"auser \& Neuh\"auser
2015c). Since it is dubious to interpret this Korean record as an aurora, it is not justified to deduce the Dst index
or storm strength from this observation -- it does not even indicate a storm. There is no evidence for an exceptional
solar super-flare, e.g. the fact that no sunspot was recorded for that time, in particular no evidence for the
2nd strongest solar flare in millennia as claimed by Mekhaldi et al. (2015). Also Stephenson (2015)
concluded that the 990ies are not unusual regarding auroral frequency.

\subsection{Aurorae records from the Qing dynasty in Kawamura et al. (2016)}

Kawamura et al. (2016, henceforth K+16) searched in the \textit{Qing shi gao} (\textit{Draft History of the Qing
Dynasty}) (specified in K+16 as AD 1559-1912 or 1616-1912 (their section 1), or 1644-1912 (their section 4), in their
table 1, entries range from 1613 to 1876, plus two without years; this dynasty ruled from AD 1616-1911 according to Xu
et al. 2000); note that the standard date for the beginning of the Qing is 1644.

K+16 found 111 entries listed as {\em aurora candidates} in their table 1, of which 77 are
{\em white}. They then exclude those 26 cases, where the observation was longer than 4 days,
which could be comets. In their table 1, there is one more entry (no. 5 in AD 1618) for {\em 16 days}, 
which is not listed among the excluded ones (end of section 3.2); apparently, K+16 did not compare
their event list with comet catalogues like Ho Peng Yoke (1962) or Kronk (1999), but simply mention the comets of AD
1668 and 1680.\footnote{Also, Hayakawa et al. (2017b) list at least two well-known comets as aurora candidates:
MS\#A10 is comet C/1618 V1 as given in Ho Peng Yoke (1962) and Kronk (1999) (Hayakawa et al. 2017b: {\em the
records are maintained on the list [of aurorae] regardless of their degree of reliabilities}, end of
their section 3) and YS\#A11 (AD 1362), for which Hayakawa et al. (2017b) only write {\em we must not exclude
the possibility that this was a comet} (their section 4), but they did not consult, e.g., Ho Peng Yoke
(1962) nor Kronk (1999), where it is given as comet for 40 day with long tail, partly in the same constellations as
listed by Hayakawa et al. (2017b) for the presumable aurora.
Abbott \& Juhl (2016) also list several well-known comets in their aurora catalogue,
e.g. the comet of AD 1695 Nov, which is observed in the same constellations Crv and Hya as listed
for the presumable aurora (e.g. Kronk 1999);
the comet of AD 1529 Feb (only one observation with a long tail in Kronk 1999) is listed
in 11 entries in Abbott \& Juhl (2016), namely from AD 1529 Jan 26 to Feb 8,
as {\em white vapor(s)}, partly in the SW, up to $60^{\circ}$ long
(not an aurora, but the comet tail).
}

Among their 14 {\em primary aurorae} listed in their section 3.5 with simultaneous observation
in Fritz (1873) on {\em the same day or one day before} (K+16, their section 3.5), but not
being a comet, five are about the event AD 1770 Sep 17+18, for which the aurora detection in China itself is also
mentioned in Fritz (1873). K+16 do not mention other publications, which already presented some of these aurorae; e.g.
the Yau et al. (1995) catalogue extends to AD 1770, but includes only two of the seven different events from K+16
(until AD 1770) as aurorae (AD 1730 Feb 15 and AD 1770 Sep 17+18), because important aurora criteria are not fulfilled:
e.g. the event on AD 1754 May 8 is likely a lunar halo display close to full moon ({\em white band from SE to W}), close
to the Schwabe cycle minimum in AD 1755.2; those on AD 1853 Apr 23 and AD 1868 Oct 30 neither give northern direction
nor night-time, and are therefore dubious as aurora candidates.

Among their 14 {\em primary aurorae}, five are on the same date and one of them plus five others
do not mention night-time (K+16 section 3.5). Of course, it is not useful to study the lunar phase distribution of
events, for many of which there is no evidence that they were at night, so that their figure 3 is useless.

Among the non-primary candidates, K+16 select those five records (two on the same date) with a lunar phase between 0.0
and 0.1 or between 0.9 and 1.0 as {\em secondary candidates} (their section 3.6). Only one of
these five records is listed in Yau et al. (1995), namely the one on AD 1618 Jul 19, for which Neuh\"auser \& Neuh\"auser
(2016) noticed that it is simultaneous to the sunspot drawn by Malapert for AD 1618 Jul 7 to 19. Among those five
records, four are within the Maunder Minimum on three different dates: AD 1650, 1667, and 1679. None of them are listed
in any other East Asian aurora catalogue, possibly because none of those reports mention night-time (two {\em white
vapor} and one {\em band of white light} without direction), which could be atmospheric optical effects.

K+16 also compare their presumable aurorae with sunspots and the sunspot Schwabe phase in their figures 1 and 2;
however, since there are many non-auroral events in their sample, it is not surprising that the auroral distribution
does not follow the sunspot Schwabe cycle phase (their figure 2). There are almost no aurorae in the rising part (phase
0.2-0.4), but 17 events are seen at minimum (phase 0.0-0.1 or 0.9-1.0), including 15 white ones. 

\subsection{The period AD 581-959 (Tamazawa et al. 2017)}

First, we will discuss aurorae, then sunspots as presented by Tamazawa et al. (2017, henceforth T+17).

\begin{figure*}
{\includegraphics[angle=0,width=17cm]{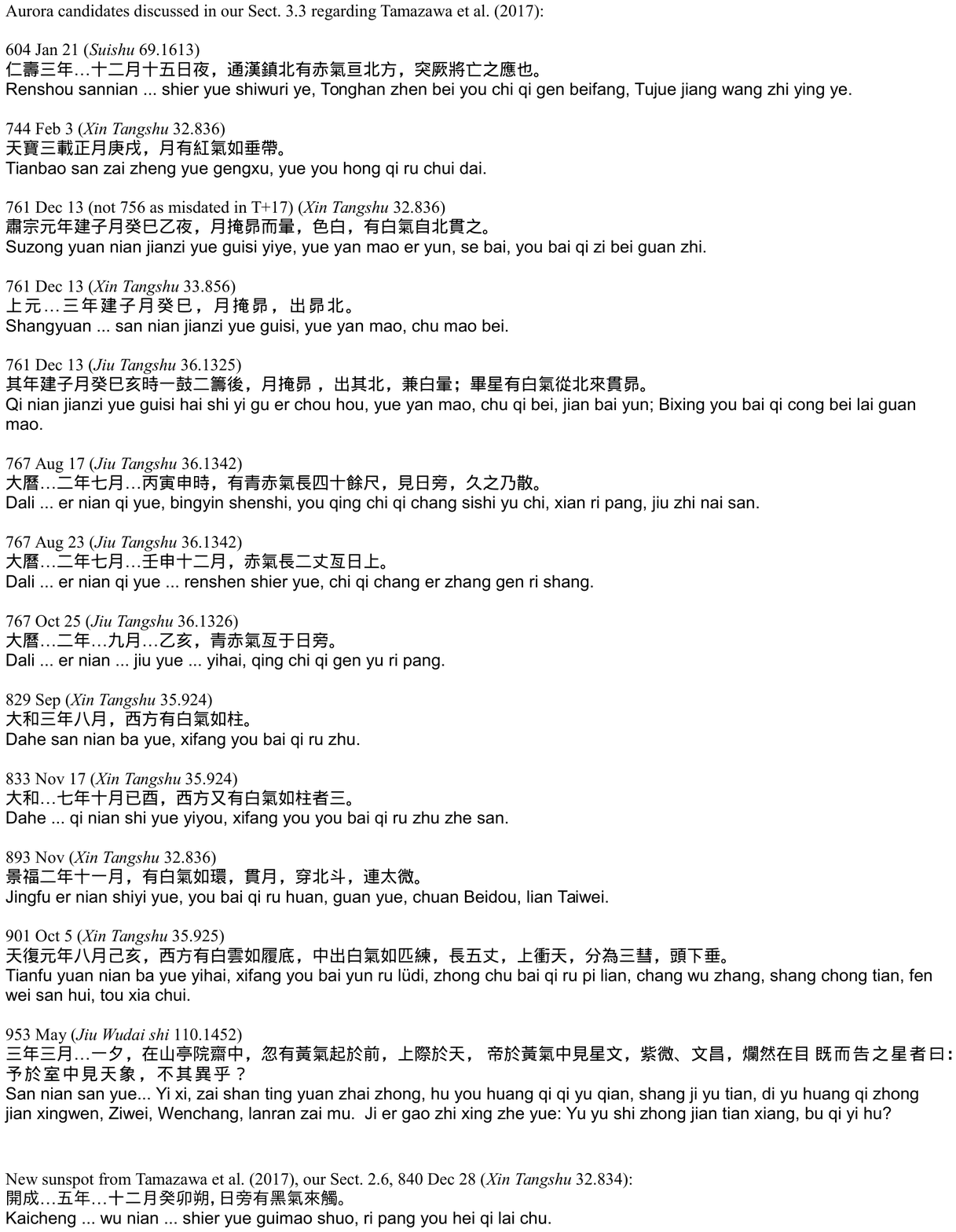}}
\caption{Here, we show the Chinese texts related to some of those events reported by Tamazawa et al. (2017),
which we discuss in our text -- first their aurora candidates, at the end the one likely true new sunspot. 
Translation and discussion are given in the Sect. 3.4.}
\end{figure*}

\subsubsection{Aurora records in Tamazawa et al. (2017)}

We will study a few aurora examples given in Tamazawa et al. (2017; T+17). Then, we present English translations
of many of their presumable aurorae (showing that several were at day-time).

T+17 used the same method as H+15 to search for aurorae and sunspots in the period AD 581 to 960 (according to their
section 2.1, but up to AD 959 according to their title; auroral records in their table 1 start in AD 511; sunspot
records in their table 4 start at AD 567). Their discussion includes records of {\em vapor, cloud, light with
color} and {\em remove(d) unsuitable ones (e.g. those observed during day-time and so on)}. 
Because it is not clear what specific criteria they used, their work is not reproducible (their
section 2.2). They then wrote that {\em Keimatsu listed all the luminous phenomena seen at night ... However,
Yau (1995) indicated that Keimatsu's work mistakenly included comets or shooting stars ... Therefore, we assume that
the records of luminous phenomena observed at night are potentially those of auroras} (their section 2.2.2).

In their example EX3 (their section 2.2.2), T+17 include the report for AD 776 Jan 12; there are three different datings
in the literature, but after a careful philological and astronomical analysis, Chapman et al. (2015) found AD 776 Jan
12 to be the only intrinsically correct one; this date is also listed in T+17 (without mentioning a reasoning). The
sequence of constellations given are those on or close to the path of the moon that night (very close to full moon), so
that Chapman et al. (2015), after a detailed discussion, considered it non-auroral and as a probable halo display (T+17
only cite \textit{Xin} \textit{Tang shu}, not \textit{Jiu Tang shu}, and do not discuss the differences in their
translation compared to several previous publications). In their section 3.4.1, T+17 acknowledge that this observation
was {\em east-south-west}. In their table 2, however, they only give
{\em west}; the translation in T+17 gives {\em in the eastern sky}.
Since the phenomenon was seen above the roughly full moon only in the first half of the night, it was only in the east
and south, see Chapman et al. 2015). They further write {\em white vapor was seen in the area whose
elevation angle was up to 40-90 degrees ... this record cannot be explained by lunar halos ... neither with a radius of
22 nor 46 degrees}; however, as seen from Chang'an (modern Xi'an), the Chinese capital at that time, the
moon itself rose up to 75 degrees above the horizon. Even the
rising moon can generate halo displays that reach the zenith. The event is not listed as an aurora in Yau et al. (1995)
nor Xu et al. (2000).

For their example EX4, T+17 also only cite \textit{Xin Tang shu}, but not the variants in \textit{Jiu Tang shu} and
\textit{Wenxian tongkao} as cited in Chapman et al. (2015), who consider it a likely true aurora like Keimatsu (1973),
Yau et al. (1995), and Xu et al. (2000), whose work again is not mentioned in connection to EX4 by T+17.

T+17 then highlight presumable quasi-simultaneous sightings of their aurora candidates elsewhere and quote in
EX6 for AD 937 Feb 14 for Europe {\em around the cockcrow continuously illuminated by day, bloody light
appeared through all of sky} (their section 3.1); while the original {\em ... usque ad
illucente die} (Odericus Vitalis in Eccl. Historia, see Link 1962) should not be translated
{\em by day}, but {\em from around cockcrow} (the first
three hours after the middle of the night) {\em and until the (first) daylight}. T+17 translate at the end of
the Chinese text for AD 937 Feb 14 (their EX5) with {\em ... and disappeared at the dusk},
while Xu et al. (2000), who bring two more reports, translate {\em ... did not disperse until
dawn} (this Chinese text shows that aurora visibility ceases at the start of twilight). The translation
by T+17 is wrong: {\em dusk} is at sunset, {\em dawn} at sunrise. The
listing for this event on AD 937 Feb 14 in their table 3 is then not consistent with their translation (vapor and
direction are missing in their table). 

T+17 list 47 aurora candidates (with {\em vapour, cloud,} or {\em light} 
in their tables 1-3), which are distributed quite equally around the lunar phase
(their figure 3). One would expect a peak around new moon, if these were mostly true aurorae, but many of their
presumable aurorae are during the day (see below), so that the lunar phase is irrelevant.

In their section 3.4.1, T+17 discuss some of the observations considered by Chapman et al. (2015) as either
misinterpreted or dubious aurorae, partly because they were reported at twilight and/or as white light near the moon.
As a counter-argument, T+17 again bring their example EX6, which ended around twilight. The second example given by
T+17 as counter-argument is a simultaneous sighting in China (T+17: {\em at sunset}) and
Japan (T+17: {\em at the beginning of night}) on AD 1363 Jul 30; Xu et al. (2000)
include in their translations for this date {\em two white rainbows surged straight up} (China,
a halo display?) and {\em as if there was a fire burning in the distance ... a sign of drought}
(Japan, a real wild fire or a deep red sunset?), so that these examples remain dubious. 

The third example given by T+17 for aurora seen at twilight is the Carrington event, which is widely considered a
particularly strong solar storm, clearly very different from the descriptions considered as misinterpreted or dubious
aurorae by Chapman et al. (2015). The arguments in T+17 in favour of historical aurorae around twilight or full moon are
misleading.

The aurora T+17 present in section 3.4.4 for AD 763 Sep (as {\em not listed in C[hapman et al.
20]15}) is not new, but has appeared in previous scholarship. This record is translated in both Chapman
et al. (2015) and Xu et al. (2000), where it is dated to AD 762 May 20. We could identify the text T+17 referred to
only by referencing their web-based catalogue as they do not give an explicit reference in their article. We are
uncertain as to how T+17 arrived at the AD 763 Sep date.

Among the 59 aurora candidates listed in tables 1-3 in T+17 (where the lunar phase is unnecessarily given to up to six
digits), ten have {\em rainbow} as textual description, four were seen towards the south, and
three are just some {\em L[ight]} (T+17 tables 1-3), i.e. highly dubious, and most others 
were previously published.\footnote{In table 5 in Hayakawa et al. (2017b), 
three of the four aurora candidates with given directions are towards
the {\em east-south}, so that they are dubious as aurora; one of them (seen for 19 days) is
comet C/1618 V1 (Ho Peng Yoke 1962, Kronk 1999).
Abbott \& Juhl (2016) also list many events in a southernly direction as aurorae,
several of them even during well-known Grand Minima -- they are all non-auroral,
but some other phenomena like e.g. clouds, halo effects, comets, etc.
}

Since it would be beyond the scope of this paper to translate and check each and every presumable aurora listed in T+17,
we present here just one case from the 9th century, which was previously published elsewhere: 

AD 811 Mar 31: T+17 list only {\em red vapor} (their table 2); Keimatsu (1974) wrote
{\em a meteor ... fell down ... There was a red vapor like a standing snake, more than ten feet (degrees)
long, upon the place where the star fell down, but it vanished in the evening ... doubtful}. This
sighting probably refers to a meteor, not an aurora, and even the {\em red vapor like a standing
snake} is reported to have {\em vanished in the evening}, showing again how
important it is to consider the context (i.e. the preceding and following sentences).

Tables 1-3 in T+17 only present certain keywords, but not the full text nor the source. The Chinese text is found only
on their webpage, but without English translation. Several of the events were clearly observed at day-time. 
All those events, which would be new (as yet unpublished) aurorae in their tables 1-3,
but which are not only about some {\em light} or {\em rainbow}, 
were translated to English by us (Chinese texts in our Fig. 2). 
Only the first event can be considered as aurora:

604 Jan 21: {\em On the night of the fifteenth day in the twelfth month of the third year of the Renshou reign
period, north of Tonghan Garrison there was crimson \textit{qi} that extended across the north. This was a response to
the imminent fall of the Tujue. (Or possibly: This was a response to the death of the Tujue general).}
The text fulfils the aurora criteria night, crimson (reddish), north; however, it was on the 15th of the lunar month,
i.e., on the night of a full moon.

744 Feb 3: {\em On the \textit{gengxu} day of the first month of the third year of the Tianbao reign
period, red \textit{qi} hung from the moon like a sash.} \\
Given the new moons on Jan 19 and Feb 18, the sighting was very close to full moon. 
The questionable phenomenon is directly related to the moon (and nothing else is
reported), so that this sighting could well be a lunar halo display, e.g. a pillar ({\em like a
sash}, i.e. a bar or belt), red perhaps due to low altitude.

761 Dec 13 (not 756 as dated in T+17): 
{\em During the second night watch on the
guisi night of the month \textit{jianzi} (the eleventh month) in the first year of Suzong (r. 756-762), the moon
occulted the Pleiades and was surrounded by a halo, white in color, and there was white \textit{qi }that came from the
north to penetrate it} (XTS 32.386).\\
The 756 date is highly doubtful. There is no \textit{guisi} [30 in the sexagenary cycle] day in the eleventh month
of AD 756, which began on a \textit{xinghai} day [48 in the sexagenary cycle]. We found two similar reports both
suggesting an AD 761 Dec 13 date. The \textit{Xin Tangshu }records {\em On the guisi day of the month
jianzi in the 3rd year of the Shangyuan reign period, the moon occulted the Pleiades. It emerged to the north of the
Pleiades. On the dingmao day of the 8th month, again it occulted the Pleiades} (XTS 33.856).
\textit{Jiu Tang shu} 36.1325 reports: {\em In that year (when the reign period name changed to Yuan), on
the guisi day of the 11th month, at the hai double hour (21-23h) after the first night drum (in) the 2nd fifth-hour,
the moon covered the Pleiades, and then emerged to its north. It was surrounded by a white halo. The stars of Bi
(Taurus) had white qi among them which followed the moon north to penetrate the Pleiades.} This text is
the most detailed among the above variants and also fully consistent as the moon did in fact occult the Pleiades in
Taurus that night, see Chapman et al. (2015) for more details on this event.

The confusion likely issues from the dating of the text in the Yuan year of Suzong. Yuan is generally the first year of
a reign period; T+17 seem to read it as {\em the first year of the reign of Suzong.} However, Yuan is in fact the name
of a very brief reign period. During the ninth month of the second year of the Shangyuan reign period, the name of
the reign period was changed to Yuan (see Fang \& Fang 1987). The eleventh month was adopted as the beginning of a
new year (see XTS 6.164). Hence, while we see the event identified as having occurred both in the Yuan year of Suzong
and in the 3rd year of the Shangyuan reign period, these refer to the same actual date. Moreover,
the text clearly refers to a halo display rather than an aurora.

767 Aug 17: {\em During the \textit{shen }double-hour (15h-17h) of the
bingyin of the seventh month of the Dali reign period, there was green/blue and crimson (qing and
chi) qi more than forty chi in length which appeared on the side of/beside the sun [ri
pang]. Only after a long while did it disperse.} \\
Here, the colored \textit{qi} {\em appeared beside the sun}, so that it is clearly some non-auroral atmospheric optical 
effect. The term {\em ri pang} can mean {\em beside the sun} or {\em on the side of the sun};
see Sect. 4.1.

767 Aug 23: {\em In the seventh month of the second year of the Dali reign period ... On the
\textit{renshen} day of the twelfth month, crimson \textit{qi} more than two \textit{zhang} in length stretched across
above the sun.} \\
The reddish ({\em crimson) \textit{qi}} was seen
{\em stretched across above the sun}, so that again it is clearly some non-auroral atmospheric
optical effect.

767 Oct 25: {\em On the \textit{yihai} day of the ninth of the second year of the
Dali reign period, green/blue and crimson (\textit{qing} and \textit{chi}) \textit{qi} stretched across on the side
of/beside the sun [ri pang].} \\
This is again clearly some non-auroral atmospheric optical effect. We
cannot understand as to why T+17 consider this and the previous two events as aurorae; they are clearly related to the Sun. 
(What is described here as {\em qi ri pang}, i.e. vapor on or next to the sun (see Sect. 4.1 for the
translation of {\em ri pang}), also cannot be interpreted as a sunspot, because it is given as green/blue and crimson and not black.)

829 Sep: {\em In the eighth month (2 Sep-1 Oct) of the third year of the Dahe reign period, in the
west there was white \textit{qi} like a pillar (or like pillars).} \\
{\em White qi like (a) pillar(s)} does not fulfil any aurora criterion.

833 Nov 17: {\em On the \textit{yiyou} day in the tenth month of the Dahe reign period, in
the west there again appeared white \textit{qi }like three pillars.}\\
Again, {\em white qi like pillars} does not fulfil any aurora criteria; night-time is again not
mentioned. {\em Three pillars} could be a halo phenomenon during sunset in the west (light
pillars of sun and two mock suns).

893 Nov: {\em In the eleventh month (12 Dec 893-10 Jan 894) of the second year of the
Jingfu reign period, there was white \textit{qi} like a ring that penetrated the moon, went through the Northern Dipper
and reached the Grand Tenuity Enclosure / Supreme Subtlety (in Leo and Virgo).} \\
The dating in T+17 to AD 893 Nov is wrong, as the 11th lunar month ran from 12 Dec 893 to 10 Jan 894. The
{\em white \textit{qi} like a ring that penetrated the moon} was obviously a lunar halo ring.
Full moon in the given period was AD 893 Dec 26 and indeed, the still almost full moon was in or just south of Leo Dec
28-30, so that its 22 deg halo ring would cross the given constellations.

901 Oct 5: {\em On the \textit{yihai} day of the eighth month of the first year of the
Tianfu reign period, in the west there were white clouds like the soles of shoes, and from among them there issued
white \textit{qi} like a bolt of silk, five \textit{zhang} in length, upwards it (or `its top') thrust toward the
heavens and separated into three comets [hui], the head dipping downward.} \\
It is not mentioned that it happened at night. Five zhang are 50 degrees. Aurora criteria are not fulfilled.

953 May: {\em In third month (16 Apr-15 May) of the third year ... One evening,
while in his study at his mountain retreat, there suddenly appeared yellow \textit{qi} before [the emperor], which rose
to reach the heavens. The emperor saw in the yellow \textit{qi} star patterns, 
Purple Forbidden (lit. Purple Palace or Purple Tenuity) 
[circumpolar stars, Chinese: Ziwei] and Wenchang [stars in Ursa Major] 
appeared brightly before his eyes.} \\
This text is not a
real astronomical record. T+17 leave out the lines that follow: {\em He then spoke to the court astrologers saying, `Isn't
it strange to see celestial images while sitting inside my room?'} (Chinese text in our Fig. 2). This again shows
that it is necessary to consider the context.

We see that several of the events listed as aurorae (or candidates?) in T+17 are clearly during the day and connected
with the Sun, while T+17 wrote in their section 2.2 that they had {\em remove(d) unsuitable ones (e.g. those
observed during day-time and so on)}. In their section 3.1, they write: {\em Lists of aurora
candidates are given in tables 1, 2, and 3 ... we found ... 45 aurora candidates (vapor/cloud/light during the
night)}. Their tables 1-3 list 59 candidates. One could consider that the 45 (of the 59) are those where
night is mentioned. They give the lunar phase for all events for which they provide an exact date. While the lunar phase
is only relevant for events occurring at night, T+17 needlessly include it for those during the day, such as the three
events in AD 767 Oct (see our translation above).

While tables 1-3 in T+17 suggest that there were all aurorae, we recommend not to use or plot these tables: among the 11
events just listed, presumable new aurorae from T+17, at most one of them can be considered a likely true aurora (AD
604 Jan 21). Given the many non-auroral phenomena in their list of aurora candidates, it is no surprise that T+17
conclude that {\em it is difficult to see the relation between long-term change of the solar activity and
records of aurora and sunspot candidates for 6-10th century} (their section 3.6), while Neuh\"auser \&
Neuh\"auser (2015a) after a careful selection of the likely true aurorae could even reconstruct the Schwabe cycle (and
more) from the mid 8th to the mid 9th century.

\subsubsection{Sunspot records in Tamazawa et al. (2017)}

We will now consider the sunspots in Tamazawa et al. (2017; T+17). T+17 reported that they {\em found 16 sunspot candidates (black spots/vapor and unusual ra[i]nbow etc. in the Sun)} (their section 3.1);
{\em 19 sunspots records}, according to their section 3.5, 16 are listed in their table 4. The
dearth of Chinese sunspots between AD 579 and 826 was of course well-known since long ago (e.g. Eddy 1976). 

All but one of the sunspots in T+17 were listed before, 
e.g. in Keimatsu (1974), Yau \& Stephenson (1988), or Xu et al. (2000). A comparison
between T+17 on the one side and Keimatsu (1974) and Xu et al. (2000) on the other side shows the following
differences: \\
567 Dec 10 in T+17, but Dec 10-18 in Xu et al., \\
577 Dec in T+17, but Dec 30 in Xu et al., \\
579 Apr 3-6 in Xu et al. missing in T+17, \\
826 May in T+17, but May 24 in Xu et al. (black spot), \\
829 Jan 20 in Keimatsu (1974), but missing in T+17, \\
832 Apr (one without day, one on Apr 24) in T+17, but only Apr 21 in Xu et al., \\
837 Dec 22 in T+17, but Dec 22-24 in Xu et al., \\
840 Dec 28 in T+17 missing in Xu et al. (see below), \\
851 Dec 2 in Xu et al. (in Japan) missing in T+17, \\
865 Feb (without days) in T+17, but Jan 31 to Feb 28 in Xu et al., \\
874 (without day nor month) in T+17, but Jan 22 to Feb 9 in Xu et al., \\
875 (without day nor month) in T+17, but 875 Feb 10 to 876 Jan 29 in Xu et al., \\
925 Dec (without day) in T+17, but Dec 29 in Xu et al., \\
927 Mar 9 in Xu et al. missing in T+17, and \\
947 Mar 3 in Xu et al. missing in T+17.

We see again, as in H+15, that the information in previous studies such as Xu et al. is much more complete. Six
sunspot records are missing in T+17; nine have a more precise date in Xu et al. 

T+17 do not mention that all but one of their records were published and translated earlier; nor did they compare the
translations. They do not give the English translation for the only new record, namely for AD 840 Dec 28 (table 4 says
just {\em BV} for black vapor). The original Chinese is shown in our Fig. 2. Our English
translation is: \\
{\em In the fifth year of the Kaicheng reign period ... on the \textit{guimao }day, the first of the twelfth
month [AD 840 Dec 28], there was black \textit{qi} on the side of the sun [ri pang] that went back and forth jostling
about.} (On the meaning of {\em ri pang}, see Sect. 4.1).

\section{General remarks on terminology}

We will now discuss three general topics from the different publications mentioned, first the meaning of the term
{\em ri pang} in sunspot and aurora records (4.1), then aurora colors in historical reports
(4.2), and finally whether and which {\em rainbows} can be interpreted as aurorae (4.3).

\subsection{The meaning of the term {\em ri pang}}

We comment here only on the discussion of the term {\em ri pang} and a few related issues in
Hayakawa et al. (2017b), not on the remaining parts of that publication. Hayakawa et al. (2017b) claim that
{\em ri pang} can mean only {\em near the Sun}. In their appendix I
with figure 10, they show three examples where phenomena near the Sun were described with {\em ri
pang}, possibly halo phenomena. These examples show that {\em ri pang} in
connection with the Sun can indeed mean {\em near the Sun, beside the
Sun}, and/or {\em next to the Sun}.\footnote{Tamazawa et al. (2017) present several cases with this term:
{\em green/blue and crimson (qing and chi) qi ... beside the sun (ri pang)} (see above for AD 676 Aug 17 
and Oct 25 and our Fig. 2), which they include among their aurorae.}
Hayakawa et al. (2017b) also acknowledge {\em black spots/vapors near the Sun as potential
sunspots}, but they {\em did not include those} with {\em ri pang} in their listing (their appendix I);
Tamazawa et al. (2017) interpret the case of AD 840 with {\em ri pang} as a sunspot (see Sect. 3.3).
Can {\em ri pang} -- depending on context -- realy mean either
{\em next to the Sun} or {\em on the/one side of the Sun}? 
The term {\em ri pang} was translated {\em on the side of the Sun} by
Wittmann \& Xu (1987), Yau \& Stephenson (1988), and Xu et al. (2000) in several cases and then
included in their sunspot catalogues. Hayakawa et al. (2017b) include \textit{Ming shi} record MS\#S24 in their list of
sunspots (without the term {\em ri pang}); however, an analogous report from \textit{Shenzong shilu} with very
similar wording includes {\em ri pang} (see Xu et al. 2000 for AD 1618 June 20-22). Another
record from the \textit{Songjiang Fuzhi} clearly says for AD 1618 June 22 {\em Within the sun there was a
black vapour} (Yau \& Stephenson 1988), a typical wording always interpreted as a spot on the disk of the
sun. We conclude that {\em ri pang} can indeed mean a spot on the sun.  
 
The Chinese naked-eye spot AD 1618 June 20-22 with the wording {\em ri pang} is corroborated 
by Malapert's drawings for June 21-29 (Neuh\"auser \& Neuh\"auser 2016, their figure 2),  
which is not the only example.

Hayakawa et al. (2017b) 
comment on this case that {\em Neuh\"auser \& Neuh\"auser (2016) relate this record [MS\#S24] 
with ... Malapert during 1618.06.21-29 ... Neuh\"auser \& Neuh\"auser (2016)
interpret the misdated intercalary month of MS\#S24 to intercalary fourth month to speculate this record as on
1618.06.20-22. This interpretation is arbitrary and no more than speculation in a philological view
point} (their footnote 14). First, we would like to note that it was not Neuh\"auser \& Neuh\"auser (2016)
that suggested to correct the date given as {\em intercalary 6th month} in
{\textit{Ming shi}} and {\textit{Tianwen zhi}} to {\em intercalary 4th month}, 
but Wittmann \& Xu (1987), Yau \& Stephenson (1988), and Xu et al. (2000), 
as clearly referenced in Neuh\"auser \& Neuh\"auser (2016, their section 3.2). Then, there are two more reports on
this spot: {\textit{Shenzong shilu}} clearly gives {\em intercalary 4th month} for 
{\em three days until day wuzi}
(June 22), see Yau \& Stephenson (1988) and Xu et al. (2000), and {\textit{Songjiang Fuzhi}}
gives {\em 5th month, 1st day}, i.e. June 22
(Yau \& Stephenson 1988). These additional reports clearly support the date correction of the
{\textit{Ming shi}} and {\textit{Tianwen zhi}} records. (Hayakawa et
al. (2017b) argue that when two spot records give {\em totally different} wordings, they
{\em can hardly be regarded as from the same origin} (their footnote 14), but we would like to
point out that two records can of course be independent from each other, and then confirming each other.
Hayakawa et al. (2017b) do not accept the date correction, they also omit the duration of three days in their MS\#S24.) 

Furthermore, Hayakawa et al. (2017b) argue that {\em the date wuzi does not
exist in the intercalary 4th month of this year} (their footnote 14), but the Chinese sources did not
specify that day wuzi would be in the intercalary 4th month. They instead say: {\em ... from intercalary 4th
month day bingxu (June 20). For three days until day wuzi ...} (June 22) in
{\textit{Shenzong shilu}} and {\em year wuwu (46th), 5th month, 1st day ...} (June 22) in {\textit{Songjiang
Fuzhi}}, day wuzi is the 1st day of the 5th month, and the spot was seen from the
2nd-to-last day in the intercalary 4th month to the 1st day in the 5th month. In their translation of sunspot MS\#S24,
Hayakawa et al. (2017b) leave out the wording {\em From day bingxu (23) until day wuzi (25)},
as given in Xu et al. (2000) from Ming shi, Tianwen zhi, san chap. 27. 
This example again shows how important it is to include all records available
(including local and inofficial ones), and not limiting a study to those available in electronic form (as done by H+15,
H+16, K+16, T+17, and Hayakawa et al. 2017b). They ignore the critical philological and historical work by previous
scholars, e.g. regarding date corrections (see appendix I in Hayakawa et al. 2017b). 
While Abbott \& Juhl (2016) dated the 1618 June 20-22 spot as, e.g., Xu et al. (2000), 
they gave intrinsically inconsistent translations:
for the one on 1618 June 20 though 22 from Mingshi Tianwen, they translate {\em ri pang} as {\em On one side of the Sun},
while in an additional record for 1618 June 22, they translate {\em qi ri pang} as {\em vapor next to the sun};
the latter record given in Abbott \& Juhl (2016) includes intrinsically that a spot was seen also June 20 and 21.
An extra sentence given in the translation of Abbott \& Juhl (2016) about the observation on
June 22 clearly shows that this was not something outside or next to the disk of the sun
(like a mock sun), but on the disk of the sun
i.e. a sunspot: {\em On the 22nd [June], at 1-3 pm, I used a bowl of water at home to observe the sun
and saw a black vapor next to the sun [ri pang]} (the correct translation here would have to be 
{\em ... black vapor/qi on the/one side of the Sun}) --
such a bowl of water as observing tool would not be needed for some black cloud
next to the sun, but only for some feature on the disk of the sun.
With the wording {\em next to the sun}, one could possibly also mean {\em before the sun},
e.g. as a shadow of a transiting small body, a theory also discussed in Europe that time.
The sunspot catalogue of Abbott \& Juhl (2016) contains only some 15 new records,
all others were given by one to several previous publications by others (mostly not cited by
Abbott \& Juhl 2016), and they missed several East Asian sunspots listed by others.\footnote{Hayakawa et al. 
(2017b) write {\em We found a probable
coincidence between Chinese sunspot records and western sunspot drawings in 1618.06/07, i.e. MS\#S22 and MS\#S23 and
sunspot drawings by Malapert (1633) as shown in Fig. 6 ... Although the sunspot size is intriguing, Malapert (1633)
seems not to draw shapes of sunspots but to place dots where sunspots
locate} (their section 4). It is not clear which coincidence they mean: they date sunspot MS\#S22 to AD
1618 May/June \ and date their sunspot MS\#S23 to AD 1618 June/July. In the above quoted statement they wrote
{\em western sunspot drawings in 1618.06/07} (i.e. without May); their drawing in figure 6
mentions sunspots in June and July in the caption, but shows sunspots from July only in the figure parts. While
Hayakawa et al. (2017b) claim to have found a coincidence between some Chinese naked-eye sunspots in AD 1618
May/June/July with some telescopic drawings, we would like to note that the coincidence of the Chinese
sunspot records of AD 1618 June 20-22 and the drawings by Malapert for AD 1618 June 21-29 were first noticed in
Neuh\"auser \& Neuh\"auser (2016), who wrote about this spot: {\em We estimated the spot size to be about 1600
square arc sec, so that the spot would not be visible to the naked eye (...); we conclude that Malapert did not draw
the spot size to scale here: note that all spots in this drawing are very similar, which is not typical -- since
Malapert supported the transit theory, his main interest may have been the path of the spot(s), shown with correct
curvature (B angle) in his drawings ...} (caption to their figure 2, which shows the sunspot drawing by
Malapert (1633) for 1618 June 21-29). Both dates in footnote 21 of Hayakawa et al. (2017b)
on sunspot MS\#S24 are wrong. Their statement
that no Western Calendar dates can be given for the 4th intercalary month (their footnote 21) is also wrong: the date
range for that month is AD 1618 May 24 to June 21, as e.g. given in their footnote 19. Also the claim by Hayakawa et
al. (2017b) that Neuh\"auser \& Neuh\"auser (2016) {\em related this record [MS\#S24] with the left
drawing of [their] Figure 6 by Malapert during 1618.06.21-29} is not correct. Figure 6 of
Hayakawa et al. (2017b) does not show the sunspots AD 1618 June 21-29 and July 7-19 as they claim in their caption
(figure 6), but the left hand-side of their figure 6 shows the sunspot AD 1618 July 7, 9, 13-15, 17, and 18 (as
observed by Malapert) and the right hand-side figure shows the spot seen AD 1618 July 8, 13, 18, and 19 as observed by
Simon Perovius SJ from Kalisz in Poland -- with dates, observers, and locations given are in the drawing and Malapert's
caption (partly also seen in the scan published by Hayakawa et al. 2017b) as well as in Neuh\"auser \& Neuh\"auser (2016).}

The sunspot of AD 1618 June/July (MS\#S23 in Hayakawa et al. 2017b), which they date to sometime from June 22 to July 21
(their footnote 20), is almost certainly the spot(s) dated to June 22 by Yau \& Stephenson (1988), would then not be
simultaneous with the aurora on AD 1618 July 19 as claimed by Hayakawa et al. (2017b, their last section); instead,
that aurora is simultaneous with the telescopic sunspot(s) of Malapert and/or Perovius AD 1618 July 7-19 (table 6 in
Neuh\"auser \& Neuh\"auser 2016), also seen in figure 6 of Hayakawa et al. (2017b), which has an inaccurate caption. It is
therefore also not justified to conclude that MS\#S23, which they cannot date well, {\em caused a flare and
brought a magnetic storm with low latitude aurora on 1618 July 19 and scientifically detected as a nitrate signal in
1619}, as Hayakawa et al. (2017b) claim. 

\subsection{Aurora color (e.g. Hayakawa et al. 2015, 2016)}

In their example quotation (a) for aurorae, listed in their section on {\em auroral records}, H+15 present an event as
{\em pale-white}, while it is given as {\em B[lue-]W[hite]} in their table 2; this was {\em B[lue-]W[hite]} in Keimatsu
(1974), {\em green-white} in Yau et al. (1995), and {\em greenish-white} in Xu et al. (2000). 

H+15 should have discussed the differences in the translations. Instead, H+15 consider the colors of \textit{wuxing}
(five elements) given by them in their section on {\em auroral records} as {\em white, red, green, yellow, and
black}, while in their section on {\em color of aurora}, the colors of
\textit{wuxing} are given as {\em white, red, blue, yellow, and black}. H+15 do not mention
that the color \textit{qing} is sometimes rendered as blue, sometimes as green.

Also, their sentences {\em there are more white auroras than the red auroras} and
{\em there are more red auroras than white (green) auroras} appear to be in contradiction (both
in their section on {\em color of aurora}). H+15 understand {\em green} aurorae to be represented with the
color \textit{bai}, conventionally translated as {\em white}, whereas the color \textit{qing} is conventionally
translated as {\em blue} or {\em green}. It may not be justified to consider events reported as {\em white} 
(e.g. {\em white qi}) to be {\em green aurorae}, as H+15 did. These events may not be aurorae at all. 

H+16 suggested that {\em white rainbows} could be aurora candidates (see next Sect. 4.3).
Because {\em white} would not be typical for aurorae, they try to explain, why this word can
appear in an aurora: {\em When observing a very faint aurora, ... the color can be recognized
as whitish because the target is not bright enough} (their section 4.1). T+17 make a similar claim:
{\em If the red aurora ... occupied the background, one would see the aurora as white} (their
section 3.2). Because {\em auroras are not really 'white'} (H+16), H+16 conclude that the
observers in fact meant green or yellow: {\em it is reasonable to assume that green or faint yellow auroras
were expressed as white as they do not have 'green' as a base color in their traditional concept} (H+16).
What if the court astronomer really meant {\em white} being one of their wuxing colors? The
color green is between the wuxing colors yellow and blue (if we read qing as blue rather than green), and there is some
green tending to yellow and some green tending to blue, so that yellow and blue would appear as more natural proxies
for green. 

Real green aurorae appear normally at low height; to observe them at low geo-magnetic latitude such as China, one would
need strong solar storms. When H+15, H+16, and T+17 (e.g. T+17 {\em think that at least some of the white
aurora candidates were indeed the green (white) auroras}, their section 3.2) they convert records of a
white phenomenon to a green aurora, i.e. to a strong storm. In doing so, they in fact modify the sources, thereby
overestimating solar activity. 

Their table 2 (H+15) includes many cases, where something {\em R[ed]} is reported at
{\em sunset} in the {\em west} (15 cases), which are almost certainly
real sunsets with glow, and also cases, where something {\em R[ed]} is reported at
{\em sunrise} in the {\em east} (six cases), which are almost certainly
real sunrises -- hence, a real {\em aurora} but not an {\em aurora borealis} (northern light). 
Previous aurora borealis catalogues did not include such events. Even though
the Chinese documents of course do not include all red sunsets nor all red sunrises, it is not justified to include
them as aurora borealis or even as candidates, because they could just have been exceptionally red sunrises or sunsets
due to, e.g., very high humidity, smoky conditions, or dust storms.

Aurora colors were recently also discussed in Abbott \& Juhl (2016),
namely as presumable indicator for the amount of $^{14}C$ production:
{\em we postulate that the most energetic (e.g. luminous) auroras will have a white component ...
The other colors of auroras: yellow, dark blue, violet, purple, and light blue ... are more
likely to represent auroral excitation during higher energy events} (their section 2.3);
this may appear contradictory.
They also considered so-called black aurorae:
{\em Black auroras are dark areas of aurora-like activity within a surrounding aurora
... The origins of black auroras are not fully understood} (Abbott \& Juhl 2016, their section 2.3).
However, if something reported as {\em black} (e.g. {\em black qi/vapor}) could be considered as
aurora, then one would definitely have to expect that aurora of some other (real) color
would also be reported, such as red qi; if no other colors are reported, than the
interpretation of something {\em black} as aurora is highly dubious.
This problem affects many presumable aurorae in Abbott \& Juhl (2016).

\subsection{{\em White} and {\em unusual rainbow} (Hayakawa et al. 2016)}

In H+16, it is suggested to interpret {\em white rainbows} (bai ni) and {\em unusual rainbows} (hong ni) as aurorae. While
H+16 point to the fact that in most cases, the {\em white rainbow} is reported to {\em pierce the Sun} or the moon (their
section 3), which they consider as {\em solar/lunar halos or similar atmospheric optics
phenomenal} (their section 3). Obviously, this is the halo phenomenon called horizontal arc, which is
white; as an arc (or bow), it can {\em pierce the Sun} or moon. H+16 found eleven cases of {\em white rainbow}
and two cases of {\em unusual rainbow}, without mention of Sun or moon, including 3 reports in the
\textit{Jin tang shu} and ten cases in the \textit{Xin tang shu}, the two official chronicles of the Tang dynasty (AD
618-907).

Other {\em white rainbows} (bai ni), not mentioned to pierce the Sun or moon, are listed under a
section heading called {\em unusual rainbow} (hong ni) in the omenology treatise
\textit{wuxing} of the \textit{Xin tang shu}. These are probably some other phenomena, i.e. not horizontal arcs through
Sun or moon. The term {\em hong ni} (translated as
{\em unusual rainbows} in H+16) is a binome composed of two different single-syllable words
both meaning {\em rainbow}. \textit{Ni}, or \textit{ci ni}, sometimes refers to a fainter,
secondary rainbow.\footnote{Chapman et al. (2015) wrote: {\em There are several different words conventionally
translated as \textit{rainbow} in classical Chinese. \textit{Daidong} appears in the \textit{Classic of Poetry}, while
both \textit{ni} and \textit{hong} appear in astronomic and omenological texts, where they are largely used
synonymously. A distinction is sometimes drawn in the genders of the \textit{ni} and the \textit{hong}; the term
\textit{ni} often occurs in the phrase \textit{ci ni}, a female rainbow. Unlike the English word \textit{rainbow},
\textit{hong} is composed of a single element, not two independently meaningful morphemes, \textit{rain} and
\textit{bow}.}} However, it is not justified to consider those phenomena as
{\em different from the atmospheric optics events of solar or lunar origin} and instead to
consider them as aurorae (H+16, their section 4.1). H+16 neglect the large variety of other somewhat significant
{\em rainbows}, which were to be reported and later compiled, e.g. fog bows, sometimes also
called {\em white rainbows}, double and abnormal rainbows, horizontal arcs away from Sun or
moon, and other halo arcs, see Minnaert (1993). Such bows and arcs were also called
{\em rainbows} in European sources until the Early Modern Period\footnote{In medieval to late 
renaissance European texts, the word usually translated as {\em rainbow} often stands for
an arc or bow from a halo display (Neuh\"auser \& Neuh\"auser 2015d); in antiquity, {\em arcus pluvius} was used for {\em rain
bow} in today's sense, {\em arcus imbrifer} for {\em rain bringing arc} (probably the $22^{\circ}$ halo ring), and {\em arcus
caelestis} for {\em arc on sky} (probably some particular halo bow); hence, there is a clear possibility to
differentiate between different bows, e.g. a real {\em rainbow} in the current modern sense which happens after a rain
around the anti-solar point or a halo arc or bow around the Sun called a {\em rain bringing arc}, which is connected to
the ancient wisdom that halo rings (or bows) are often followed by rain, an incoming depression (e.g. in
{\em Meteorology} by Aristotle), see also Minnaert (1993).} and
they all, whether day or night, are related to Sun or moon due to reflection or refraction. The statement by H+16 that
{\em auroras appear at night, and the usual rainbows appear during the day-time} (their section 3) is misleading.

At least seven of the eleven {\em white rainbows} (listed in H+16) were observed westwards
({\em laying across westwards} or similar). A white bow in the west does not fulfil any of the
aurora criteria from Neuh\"auser \& Neuh\"auser (2015a). Only in two of those cases, it was reported that it happened at
night, namely JT 2 and XT 10. (H+16 claim (at the end of their section 3 after the translations) that XT 10 would give
night, but that it is not given in their translation; however, the word {\em night} is indeed there in the Chinese text.)
Observations of phenomena which are described in similar form for day- and night-time, are not good candidates for rare
auroral displays which can be detected only at night. 

In XT 5, a {\em white rainbow} is reported to {\em lay across east to west} (H+16), 
not atypical for aurorae. It is not justified to conclude that the description
{\em filled up the heaven} would mean to be {\em spreading across the sky through the
region around the zenith} and {\em likely to imply aurora}, as done by H+16,
given that they neither know what kind of {\em white rainbow} is meant nor what the given size
of {\em as wide as 5 chi} means (see also the end of their section 3); {\em 5 chi} are some 5 degrees, 
consistent with the width of a fog bow (Minnaert 1993). JT 3 and XT 2 have the wording
{\em an unusual rainbow filled up the heaven/sky}. Again, a phenomenon called rainbow, a
somewhat narrow band, cannot really fill up the whole sky as implied by H+16, but one could use this wording to mean
something crossing large areas of the sky with a large angular extend, e.g. all sorts of bows mentioned above. All
three reports (JT 3, XT 2, XT 5) do not mention night-time. In both XT 2 and example (a) given in section 4.2 in H+16
(AD 1363 July 30: {\em two white rainbows, which directly hit the Plough}), H+16 incorrectly translated
the Chinese {\em Beidou} to {\em Plough}, one should use {\em Northern Dipper} or
{\em Big Dipper}, the constellation referred to here. 

According to two more records, a white rainbow {\em descended into the city} (XT 1) and
{\em descended to the gate of their camp} (XT 3), so that these phenomena seem to have been observed nearby and low; the
impression of closeness is 
atypical 
for aurorae, but often reported for halo displays as well as fogbows and rainbows.
As these two events coincide with military events (normally at day-time) in the locations where they occur, the records
should be regarded as somewhat suspect.

Xu et al. (2000), the most recent critical compilation of East Asian aurorae for the last two millennia, excludes all
records which do not explicitly report night-time. The claim by H+16 that {\em there are no explicit records
of day-time observations of 'white rainbows' or 'unusual rainbows' in our list} is again strongly
misleading. Obviously, the professional Chinese court astronomers did not explicitly specify whether something was at
day- or night-time, when it was during the day-time -- there may well be a few rare exceptions (e.g. SN 1054).
Therefore, when H+16 imply a {\em lack of information} regarding day- or night-time, they
oversee that the information is implicitly given. 

The listing in H+16 includes only one case mentioning more than one band, namely XT 4, the {\em four white rainbows ...
observed at night} on AD 757 Feb 20. While it may in principle be possible to observe several atmospheric optics arcs
or bows at night at once (moon-related), this observation was close to new moon, which is the main argument for
considering it seriously as aurora candidate. Among all the 13 {\em white rainbows} or so-called
{\em unusual rainbows} presented in H+16, only the one in AD 757 (XT 4) could be auroral, as
considered by others.\footnote{H+16 wrote that {\em previously, white rainbow was not considered to be an aurora}
and that {\em Chapman et al. (2015) also mentioned a possibility that white rainbow in 757 could perhaps be an aurora}
(H+16, their section 1). Chapman et al. (2015) in fact considered the report about {\em four white rainbows} in AD 757 in
their section on {\em questionable aurora reports} and wrote that this event was {\em listed as an aurora in Keimatsu (1973)
("probable to doubtful"), Yau et al. (1995), and Xu et al. (2000)}.}

Then, T+17 also include several cases of {\em rainbow} in their list of (presumable) aurorae,
all or most of which may well be some atmospheric optical phenomena (see our Sect. 3.3.1). Given that aurora colors and
forms might in some historical cases have been compared to rainbows (e.g. {\em like a
rainbow}), Carrasco et al. (2017) also searched for such reports; since such reports would always remain
somewhat dubious, they can never be strong aurora candidates.\footnote{The three unusual rainbows presented in
Carrasco et al. (2017) as possibly auroral could also be other phenomena: in particular their 2nd case, seen in Rio,
Brazil, where a reknowned scientific scholar (Sanches Dorta) reports {\em a rainbow produced
by refraction of the rays of the moon ... bases on the WSW and ESE ... very white}, one should consider
an arc of the white horizontal lunar halo circle -- halo bows were often called {\em rainbow}
in former times (see previous footnote) -- 
here seen in the evening (6h10m to 7h18m) formed by the moon moving from the N to the W, due to
ice crystals from an incoming depression (later {\em white cloud} in the S and
{\em lighting} in the ESE), at least high humidity ({\em whole atmosphere was
slightly reddish}). Although some aurora criteria seem to be fulfilled, the whole report is more
consistent with something else (see introduction to this paper and as already stressed in Neuh\"auser \& Neuh\"auser
2015a); the horizontal halo arc can be seen up to 360 degrees all around the horizon, or in parts (arcs/bows), and it
can change in extend and intensity according to weather conditions and illumination; if this event would have been an
aurora, seen at the very low latitude of Rio, it would have been a very strong solar storm, but there are only few
other reports those days (at mid-latitude only). Their 3rd case was {\em a perfect rainbow ... like
ashy} at full moon around the time of a {\em small storm}, seen for a few minutes
in the north, which could well be a fogbow or lunar rainbow (see Minnaert 1993). Only their first case is during the
{\em dark night}, i.e. without the moon, a feature between N and W, which changes its shape,
described {\em as fire swords ... band or circle of fire as a rainbow}. This could be an
aurora.}

\section{Conclusion}

We have critically reviewed the articles by Hayakawa et al. (2015, 2016, 2017ab), Kawamura et al. (2016), and Tamazawa
et al. (2017), where they searched electronically for certain words presumably referring to aurorae and sunspots. Apart
from internal inconsistencies within the papers, unclear or missing criteria, translation errors, omission of relevant
details, and the fact that they do not cite most of the other articles which presented their likely true spots and
aurorae before,\footnote{
This is similar to the study of Hayakawa et al. (2017c), and we otherwise do not examine it.
The only sunspot presented (7th-9th century Japan) was already known: Hayakawa et al. (2017c) refer to this record
as an instance of \textit{ hei dian} as a \textit{ black spot}, while both Wittmann \& Xu (1988) and Xu et al. (2000)
translate it as \textit{ black dot}. Hayakawa et al. (2017c) then list 13 presumable aurorae.
With one exception, these either have been published previously or refer to non-auroral phenomena,
including several instances that occur during the day or near the full moon;
their figure 4 with one or two presumable aurorae per lunar phase bin is all but meaningless.
In detail, the 13 presumable aurorae have the following issues:
(i) The sightings for 20 Dec 620 (their A1) and (ii) 18 Sep 682 (A2) were listed
in Yau et al. and Keimatsu et al., respectively.
(iii) The aurora dated by Hayakawa et al. (2017c) as \textit{ A3: 08 Oct 839 ... Translation: On 10 August 839 ...}
was dated AD 839 Aug 10 in, e.g., Yau et al. (1995) and Xu et al. (2000), where the translation is also
significantly different. (iv) The presumable aurora listed as \textit{ A4: 27 July 847 ... a white vapor emerged
out of the lunar halo and surrounded it} was not given in any other aurora catalog before, because it was a lunar
halo phenomenon a few days before full moon. (v) The presumable aurora listed as
\textit{ A5: 05 Nov 857 ... during the day, a white cloud
appeared ... from the east to west} was not given in any other aurora catalog before, because it was
not during night-time. (vi) The presumable aurora listed as \textit{ A6: 24 July 858 ... early in the morning, a white
cloud extended from NW to SW ...} was not given in any other aurora catalog before, because it was not during dark night.
(vii) 13 Nov 859 (A7) also in Yau et al. (1995).
(viii) The presumable aurora listed as \textit{ A8: 09 Nov 864 ... during the night, a light appeared on the northern
mountains and was as intense as lightning. Red light was also observed in front of the Suzaku Gate ...}
was not given in any other aurora catalog before; among the two phenonema reported,
the first is given with terms atypical for aurorae, it is not explicitely reported to be on sky,
our literal translation of the Japanese text is
\textit{ As for the northern mountain(s), there was light};
in the second, \textit{ Suzaku Gate} is the Japanese pronounciation for \textit{ Zhu Que}
for \textit{ Crimson Sparrow}, it may refer to a set of seven lunar lodges ranging from Well
(in Gemini) to Axletree (in Corvus), during this night, the area from Gemini to Corvus
rose roughly in the East, while the moon was far away setting in the West;
alltogether, it could have been an aurora;
this context should have been discussed in detail by Hayakawa et al. (2017c).
(ix) The presumable aurora listed as \textit{ A9: 17 July 865 ... around dawn, the color of the moon
was purely yellow. There was a red cloud covering it} was not given in any other aurora catalog before,
because it was not at night. (x) The presumable aurora listed as \textit{ A10: 28 Aug 876 ... at dusk ...}
was not given in any other aurora catalog before, because it was not at night.
(xi) The presumable aurora listed as \textit{ A11: 16 Oct 876 ... a white cloud appeared in the southern heavens,
extending across from the E to W} was not given in any other aurora catalog before, because it was
an extended \textit{ white cloud} in the south, close in time to full moon.
(xii) The presumable aurora listed as \textit{ A12: 20 Apr 883 ... at twilight, a lunar halo came ... a white vapor
emerged from the north and entered into the halo ...} was not given in any other aurora catalog before, because it was
a lunar halo effect a few days before full moon.
(xiii) The presumable aurora listed as \textit{ A13: 14 Aug 885 ... a blue cloud appeared in the heavens, extending
from the NE to SW} was not given in any other aurora catalog before -- night-time is not mentioned and a partly
southern direction and purely blue color (without red) would also be dubious for an aurora.}
there are strong shortcomings in the digital method as applied by these authors, e.g. they missed a
number of entries found in other catalogues based on the same primary sources. Several of these publications present
most of the Chinese texts only on a web page, without English translations. They ignore critical philological and
historical work by scholars, e.g. regarding corrections of dates due to mistakes by copying scribes
(and, e.g., the meaning of {\em ri pang}, see our Sect. 4.1). 
Their arguments
are sometimes misleading (e.g. regarding aurora colors, see our Sect. 4.2). They make unjustified assumptions, e.g.
that observations without explicit mention of day- or night-time, would be at night (our Sect. 4.3). The previously
mentioned recent publications restricted the search to official Chinese dynasty reports, which limits the resulting
completeness, as there were also reports on celestial phenomena in local gazetteers etc.

Many presumable new aurorae (and sunspots) were previously (and rightfully) rejected by others, because they are much
more likely other 
celestial or atmospheric events,\footnote{
This is indirectly confirmed by Kataoka et al. (2017), who
investigated presumable aurora candidates in H+15 and T+17, which happened within 10 days,
i.e. a small subset, they wrote:
\textit{ We investigated the occurence pattern ... of historical auroras by Tamazawa et al. (2017) and Hayakawa et al. (2015).
Their terms in oriental historical documents are known as "vapor", "cloud", "light", etc. ... a detailed visual
inspection of the records was conducted to eliminate meteorological phenomena, such as sundog-like events}
(their section 3). This means that halo (and other non-auroral) effects are still listed among the presumable aurorae in
H+15 and T+17.}
like particularly red sunrises and sunsets or well-known comets; many of the
{\it white} and {\it unusual rainbows} could easily be fog bows or other atmospheric optical phenomena like in particular
halo phenomena. H+15, H+16, K+16, Hayakawa et al. (2017ab), and T+17 
do not discuss such differences in classification compared to
previous publications. In their event tables (partly inconsistent with their figures, partly with wrong datings, and
including some duplications), they present only a few abbreviated keywords from the texts, they do not indicate whether
night-time was specified for presumable aurorae (but they give lunar phases also for day-time observations which mention
the sun), so that the reader cannot consider, whether alternative interpretations are more probable. We therefore
cannot recommend their lists for solar activity reconstructions. Moreover, among eleven presumable new aurorae from
Tamazawa et al. (2017), for which we present the original Chinese and a first English translation, at most one of them
might be a true aurora; some others being halo phenomena (partly mentioned to be located beside the sun). There have
also been other cases in which halo effects were misinterpreted.\footnote{(i) Zolotova \& Ponyavin (2016) recently
suggested an event in the deep part of the Maunder Minimum to be an aurora, which was found to be a halo display by
Usoskin et al. (2017). (ii) Usoskin et al. (2013) considered three events in the mid 770ies as aurorae (two of them also
listed in Link 1962), which were in fact halo displays (Neuh\"auser \& Neuh\"auser 2015b). 
(iii) Allen (2012) performed {\em a quick
Google search} and presented a {\em red crucifix after sunset} in the {\em Anglo-Saxon Chronicle}, presumably
observed in AD 774, as absorbed supernova in the context of a strong radiocarbon variation around AD 775
(suggested as aurora by, e.g., Usoskin et al. 2013); however, neither Allen (2012) nor Usoskin et al. (2013)
took into account that critical editions by scholars of Anglo-Saxon history date this entry to AD 776 (e.g.
Whitelock 1979), nor that it was classified (and dated) correctly as a halo display (in AD 776) by Newton (1972).
This {\em red cross} received much scholarly and public attention and was misinterpreted also as 
airglow etc.; see Neuh\"auser \& Neuh\"auser (2015b) for a detailed discussion.
(iv) Abbott \& Juhl (2016) list $\sim 150$ entries in their aurora catalogue only from AD 1523 to 1566,
which includes the Sp\"orer Minimum.
All these records are from Korea, where (like in other East Asian countries) 
both weather and astronomical phenomena are recorded together.
Most of these presumable aurorae are either white only or black only,
so that they could easily be non-auroral meteorological phenomena like halos, often even named explicitely as such,
and some are comets (see, e.g. footnote 2).
E.g., for AD 1526 Mar 22, they give: {\em Night. The moon haloed and there was a white vapor passing
through it transversely} (only 5 days before full moon).
Also, if there is a halo during the day and then some {\em white} or {\em black vapor} during the next night, which does not fulfil
any more aurora criteria (than night-time), then it can very easily be a lunar halo effect, given a similar weather pattern
during day and night, namely an incoming depression (sometimes thunder and lightning are also mentioned). 
Abbott \& Juhl (2016) specify some terms (to be found in their aurora catalogue) in
their appendix A for table A4: {\em (2) Embracement (semi-circles / found by the side of and concave towards the sun},
and {\em (4) Opposition (a bluish-white and red vapor shaped like the new moon but convex towards the sun},
which are obviously both observed during the day, i.e. non-auroral, but certain halo effects;
and {\em (3) Hong (daytime: parhelic circle/arc, nighttime: arc)}, 
but there is no reason to assume that {\em hong} cannot mean a lunar halo effect.
In their aurora table to start at AD 500 (supplementary material), they also include the {\em red cross}
in AD 774 and the {\em two burning shields} in AD 776, which are both non-auroral (Neuh\"auser \& Neuh\"auser 2015bd).
Given that very many non-auroral phenomena are included in their aurora catalogue,
all the statistics and conclusions on aurora color distribution in Abbott \& Juhl (2016) are unjustified,
and it is not a surprise that Abbott \& Juhl (2016) have to conclude:
{\em The results for blue and violet/purple
auroras are inconsistent with their known physics} (their section 5).
}

We therefore recommend great care when applying the automatic search technique to digitally available data bases to find certain
phenomena. To study solar activity, one does not need as many aurorae or sunspots as possible, but credible ones.
Electronic searches are a valuable tool. Yet they should always be applied to the full relevant corpus and be
employed in conjunction with previous scholarship. Moreover, to understand historical reports, a critical historical
exegesis has to be applied considering context, terminology, and dating. Any such search must by unbiased and
open-minded for celestial {\em signs} other than sunspots and aurorae, namely considering the vast variety of
astronomical and meteorological phenomena. 

\acknowledgements
We obtained data on lunar phases from Rita Gautschy at
www.gautschy.ch/\~{}rita/archast/mond/Babylonerste.txt (Gautschy 2014). 
We would like to thank Daniela Luge (U Jena) for help with the translation of the Latin text from
Odericus Vitalis.

\end{document}